\documentclass[11pt]{article}  
\usepackage{epsfig}

\newtheorem{theo}{Theorem}  
  
\newtheorem{proposition}{Proposition}  
\newtheorem{corollary}{Corollary}  
\newtheorem{lemma}{Lemma}  
  
\newtheorem{example}{Example}  
\newcommand{\ignore}[1]{}

\def\proof{\noindent {\bf Proof : }}

\newcommand{\Odd}{\mbox{{\sc Quad}}}  
\newcommand{\Mark}{\mbox{{\sc Mark}}}  
\newcommand{\Max}{\mbox{{\sc Max}}}  
\newcommand{\Power}{\mbox{{\sc Exp}}} 
\newcommand{\Exp}{\mbox{{\sc Exp}}}
\newcommand{\Expw}{\mbox{{\sc E}}}

\newcommand{\border} {\mbox{tail}}  
\newcommand{\stale} {\mbox{idle}}  
\newcommand{\cat} {\mbox{cat}}

\newcommand{\stato} {\mbox{{\em state}}}  
  
\newcommand{\dx} {\mbox{{\em right}}}  
\newcommand{\sx} {\mbox{{\em left}}}

\newcommand{\qed}{\hfill\rule{2mm}{2mm}}

\renewcommand{\thefootnote}{\alph{footnote}}

\begin{document}  
  
\date{}  
\author{{\large 
J. Gruska\footnote{
Faculty of Informatics, Masaryk University, Brno, Czech Republic.}
\hspace{1.0 cm}  
S. La Torre\footnote{
Facolt\`a di Scienze Matematiche, Fisiche e Naturali,
Universit\`{a} degli Studi di Salerno, Baronissi, 84081, Italia.} 
\hspace{1.0 cm}  
M. Napoli$^\ddag$
\hspace{1.0 cm}  
M. Parente$^\ddag$}}
\title{ 
{Various Solutions for the Firing Squad Synchronization
Problem.}
\footnote{Work partially supported by the grant
``{\em Metodi Formali ed Algoritmi per la Verifica di Sistemi Distribuiti}'', 
Universit\`a degli Studi di Salerno. The first author is
also supported by the grant GA\v CR, 201/04/1153.
}
} 
\thefootnote    
\maketitle  

\begin{abstract}
We present different classes of solutions to
the Firing Squad Synchronization Problem on networks of
different shapes. The nodes are finite state processors that
work at unison discrete steps. The
networks considered are the line, the ring and the square.
For all of these models we have considered one and two-way
communication modes and also constrained the quantity of information
that adjacent processors can exchange each step.
We are given a particular time expressed
as a function of the number of nodes of the network, $f(n)$ and 
present synchronization algorithms in time $n^2$, $n \log n$, $n\sqrt n$,
$2^n$. The solutions are presented as {\em signals} that are used
as building blocks to compose new solutions for all times expressed
by polynomials with nonnegative coefficients.
\end{abstract}

\section{Introduction}
The famous firing squad synchronization problem (FSSP), is an old problem
posed by Myhill in 1957 (in print in \cite{Mo64}). In terms
of Cellular Automata, we are given a line of $n$ identical cells
(finite state machines) that work synchronously at discrete time steps,
initially a distinguished cell (the so called {\em general}) 
starts computing while all others are in a quiescent state;
at each time step any cell sends/receives
to/from its neighbours some information about their state at the preceding
time: the problem is to let all cells in the line enter
the same state, called {\em firing}, for
the first time and at the very same instant, the {\em firing time}.

In literature many solutions to the original problem and to some
variations of it have been given. 
The early results all focused on the synchronization in minimal time: 
Minsky in~\cite{Mi67} showed that a solution to the FSSP requires at least $2n-1$ time,  
Waksman~\cite{Wa66} and Balzer~\cite{Ba67} gave the first solution
in this minimal time
and Mazoyer in~\cite{Ma87} constructed a minimal time solution
with the least number of states to date: six.
In \cite{Ba67} it has also been shown that five states are
always necessary for a solution.

A significant amount of papers have also dealt with 
some variations of the FSSP. These variations concerned 
both the geometry of the network and 
some computational constraints.
In the following we briefly recall some of them.
The FSSP has been studied 
on a (one-way) ring of $n$ processors \cite{Cu89,LNP97}, 
on arrays of two and three-dimensions \cite{Sh74,Ko77}: in all
these papers all the results focused
on lower and upper bounds on the minimal time for the
synchronization. In the very recent paper \cite{Ko01} the cells of
the network are placed along a path in the two-dimensional array space, there a
combinatorial problem (for which only exponential algorithms
are known) is reduced to the existence of an optimal solution to the FSSP
on this path.
In~\cite{Ro95} solutions for the Cayley graphs are given
and in~\cite{NH81} a particular class of graphs is studied
and for this class a solution in time $3r+1$ or $3r$ is given,
where $r$ is the longest distance between the general and
any other node (the radius) of the graph.
Some constrained variants of the FSSP have concerned solutions on
the interesting model of reversible CA (i.e., backward deterministic CA) \cite{IM96} and
CA with a number-conserving property (i.e., 
a state is a tuple of positive integers whose sum is 
constant during the computation) \cite{IMS98}.
Other kinds of constraints which have been considered
concern the amount of information 
exchanged between any pair of adjacent cells. 
In \cite{Ma96,LNP98} the network is a line of 
cells that can exchange only one bit, that is
at each time step 
each cell sends/receives only one bit of information
to/from the adjacent cells instead of its whole state.
Finally let us recall the significant work of \cite{CDDS89} where
the FSSP is studied in a distributed setting (no global clock, but
lock-step synchrony) with bounds on the number of faulty processors.

In this paper we consider the problem on various networks (line, ring, square),
and for one and two-way communication modes, but with a new
approach with respect to the past: we hypothesize we are given the firing
time and we ask for a synchronization algorithm in this time.
This is an interesting and challenging theoretical problem, which
is also directly connected to the
sequential composition of cellular automata. 
Given two cellular automata $A$ and $B$ 
computing respectively the functions $f$ and $g$, 
the sequential composition of $A$ followed by $B$ is the cellular 
automaton obtained in the following way: first $A$ starts on a standard
initial configuration and when it has done with its computation,
$B$ starts using the final configuration of $A$ as initial configuration.
The resulting automaton clearly computes $g\circ f$. 
In order to compose the two automata it is necessary to synchronize 
all the cells that will be used by $B$ at the time $A$ computes 
$f$.

Some of the results presented here are a revisiting and a generalization of some
results of \cite{LNP97, LNP98, LNP00}, anyway
here we present a whole framework of {\em signals} that, informally
speaking, is a set of cells that at a given time
receive or send a particular state. We then define some basic signals
(building blocks) and give some rules to combine them to obtain
other new signals. This modular approach allows to design 
synchronizing algorithms in a very natural way
also simplifying their understanding and descriptions. 
Moreover here we introduce also as a parameter the number of 
bits that can be simultaneously transmitted at each step.
We study networks where at each step a cell can transmits to 
each of its neighbours at most $c$ bits, $c \geq 1$.

As said above
the communication between adjacent cells can be in 
both directions or only in a direction. We thus consider either 
networks where a cell can exchange information with 
all its neighbours, or networks where for each cell, 
only a predetermined half of its neighbours can send information to it while
the other half can only 
receive information from it (the information flow is 
unidirectional).
In this second case, to guarantee the communication from a
cell to all the others, we consider 
circularly shaped networks. 


For all the considered networks we prove a lower bound on 
the time of a synchronization, then we prove its tightness
by giving a matching synchronization.
We obtain families of solutions to the considered variants of 
the FSSP in several times $t(n)$, where $n$ is the number of nodes
of the network. 
The approach we follow is compositional: 
we first describe basic synchronizing algorithms 
and then we give general rules to compose synchronizations.
The basic synchronizations in turn are obtained by composing
elementary signals, which can be seen as fragments of cellular
automata.
A {\em synchronization} is thus a special signal obtained as
a composition of many simpler signals.
Compositional rules for both signals and synchronizations include
{\em parallel\/} composition, {\em sequential\/} composition, and 
{\em iterated\/} composition.
We also state some sufficient conditions to apply them.
In the parallel composition we start many synchronizations 
or signals, all at the same time. In some cases, this composition
can be used to select among different synchronizations depending 
on the number of cells in the network. 
Sequential composition appends a synchronization (or a signal)
to the end of another signal, possibly with a constant time offset. 
This way we are able to construct a synchronization in time 
$t_1(n)+t_2(n)+d$, for $d\ge 0$, if there exist synchronizations 
in time $t_1(n)$ and $t_2(n)$.
If we are given two synchronizations respectively in time 
$t_1(n)$ and $t_2(n)$, the iterated composition consists of 
iterating $t_2(n)$ times 
the synchronization in time $t_1(n)$, 
thus obtaining a new synchronization in
time $t_1(n)\cdot t_2(n)$. 
Compositions of synchronizations are used to 
determine synchronizations in a ``feasible'' 
time expressed by any polynomial with nonnegative coefficients.
Finally, 
we give a construction to ``inherit'' synchronizations on 
two-dimensional networks starting from synchronizations of 
the corresponding linear networks.
We show that an $(n\times n)$ array of cells can be seen as
many lines of $(2n-1)$ cells
(each of them having as endpoints cells $(0,0)$ and $(n-1,n-1)$)
and a given synchronization on a line can be executed simultaneously on all
these lines. 
Thus we can synchronize
an $(n\times n)$ array in time $t(2n-1)$, provided that
there exists a synchronizing algorithm for a line of $k$ cells 
in time $t(k)$.

As building blocks for the compositional rules we 
give synchronizing algorithms in some common functions:
$n^2$, $n \lceil \log n\rceil$, 
$n\lceil\sqrt n\,\rceil$ and $2^n$.
To synchronize a line of $n$ cells in time $t(n)$ we first
design some basic signals and then we compose them to obtain an overall
signal that starts from the leftmost cell
and comes back to it in exactly $(t(n)-2n+1)$ time units; then
a minimal time synchronization starts, synchronizing
the $n$ cells in time $t(n)$.
To obtain a synchronization in time $t(n)$ of an array of 
$(n\times n)$ cells we use the following approach:
first synchronize a row in time $t_1(n)$ then 
start a synchronization in time $t_2(n)$ on all the columns
such that $t(n)=t_1(n)+t_2(n)$.

It is worth noticing that the composition rules also apply to the
general case of $(m\times n)$ arrays. Thus 
all the synchronizations given for an $(n\times n)$ array
can be extended to an $(m\times n)$ array, considering the time of
the synchronization as a function of either $m$ or $n$.

The remainder of this paper is organized as follows. 
In section \ref{preli} we give the definitions and introduce the 
notation we will use throughout the rest of the paper. 
In section~\ref{mts} we give tight lower bounds on the
time synchronization of $c$-CA and solutions in minimal
time. In section~\ref{sig} the framework of the signals is presented
formally. In section~\ref{composition} some composition rules on 
synchronizations are defined.
In sections~\ref{ttime} and~\ref{ttimm} solutions in the given times
$n^2$, $n \lceil \log n\rceil$,
$n\lceil\sqrt n\,\rceil$ and $2^n$  are given for the
two-way and one-way communication models, respectively.
As an application of the compositional rules to obtain 
new synchronizations, 
in section~\ref{poly} we show how to obtain 
polynomial-time synchronizations on all the considered
models.
The conclusions are in section \ref{conc}.

\section{Preliminaries}\label{preli}

In this section we give the basic definitions, introduce the models, which
are generalizations of the well known model of  cellular automata, and define
our synchronization problem.

\smallskip\noindent{\bf The models.}   A cellular automaton is an array of pair-wise connected finite-state
machines, called {\em cells} (or sometimes {\em processors}), which operate synchronously at discrete time steps. 
We consider both one-dimensional and two-dimensional cellular automata.
The connections between cells may be either one-way or two-way links.
We consider a generalization of the known cellular automata
since in our models 
the capacity of the channels, and then the communication complexity,
may vary. We call a $c$-link
a channel being able to transfer $c$ bits  simultaneously.
All the cells are indistinguishable, anyway
for descriptive reasons, in a one dimensional array of $n$ cells
we will number them starting from 0; moreover cell 0 and cell $n-1$ are 
said boundary cells.
Unless stated otherwise, in the following $n$ is the number of cells of the one-dimensional cellular automaton.
 
The behaviour of each cell is in accordance to finite state transition
functions depending on both the state of the cell and the
output given at the preceding step by some of the connected cells. We define a function
$N:\{0,\ldots,n-1\} \rightarrow \{0,\ldots,n-1\}^* $ which
determines the neighbouring
cells on which the
transition function of a given cell  depends.
This function depends on whether the connections are
one-way or two-way-links and may also vary for
different cells (for example, in the case of the boundary cells).
For a cellular automaton $A$, we denote by $m_A$  the maximum
length of $N(i)$, for $0 \leq i \leq n-1$.

\begin{figure}[tb]
\centerline{\epsfig{figure=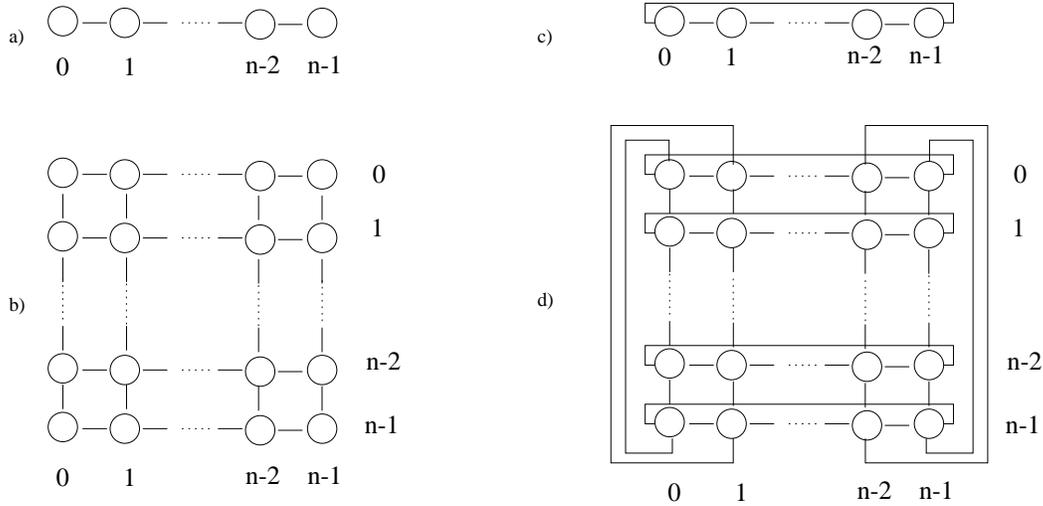, width=14cm}}
\caption{The one-dimensional and two-dimensional cellular automata.}\label{CA}
\end{figure}

A {\em c-Line} is a one dimensional cellular automaton 
where the connections are two-way $c$-links and where the $i$-th cell is
connected to the ($i-1$)-th and ($i+1$)-th cells, for $0<i<n-1$,
the first cell is connected
only to the second cell, and the last cell is connected only to the $(n-2)$-th cell,
thus $N(i)=(i-1,i+1)$, for $0<i<n-1$,
 $ N(0)=(1)$ and  $ N(n-1)=(n-2)$ (see Figure~\ref{CA}.a).
A {\em c-Ring} is a one dimensional
cellular automaton with two-way $c$-links
with a connection also between
the first and the last cell (see Figure~\ref{CA}.c). Thus it has a circular shape
and the length of $N(i)$ is two, for every $i$.
Finally, a {\em c-ORing} is a one dimensional
cellular automaton with one-way $c$-links
such that a connection exists also between
the first and the last cells. The $c$-ORing has also a circular shape
with $N(i)=(i-1)$, for every $i>0$, and  $N(0)=(n-1)$. Thus  $m_A=1$: 
the $i$-th cell
receives only the output of the $(i-1)$-th cell.
 
The two-dimensional case is a natural generalization of the already
considered models. In a two-dimensional array
of $n \times n$ cells, the cells are numbered $(i,j)$, starting from $(0,0)$.
In what follows $n \times n $ is always the number of cells of 
the two-dimensional cellular automaton. 
Each cell $(i,j)$, except for the boundary cells, is connected
to cells ($i-1,j$), ($i,j-1$), ($i+1,j$) and ($i,j+1$).
In this case, if the connections are two-way links, then
$N(i,i)=((i-1,j), (i,j-1), (i+1,j), (i,j+1))$ and,
with one-way connections, $N(i,j)=((i-1,j), (i,j-1))$.

We consider  a {\em c-Square}, where the connections are
two-way $c$-links and each boundary cell is connected only to
the neighbouring cells (see Figure~\ref{CA}.b). 
For example, in this network the cell $(0,0)$ is connected
to the cells $(0,1)$ and $(1,0)$, while a cell $(i,0)$ is connected
to the cells $(i,1)$ and $(i-1,0)$ and $(i+1,0)$.
On the other hand, we can
define the {\em c-Square of Rings},
where, similarly to the first and last cells in the  c-Ring, the boundary
cells are pair-wise connected, and {\em c-Square of ORings} where 
the connections are one-way $c$-links
(see Figure~\ref{CA}.d).

For simplicity, we do not consider the rectangular models,
that is those obtained from arrays of $m \times n$ cells.
Many of the results in this paper can be extended to this case.
Figure~\ref{schema} summarizes the considered models with respect to
both the paradigms non-circular vs. circular and two-way vs. one-way links.
Observe that we do not consider the non-circular models with
one-way links. These models are not meaningful in this context.

\begin{figure}[tb]
\centerline{
\begin{tabular}{|l|c|c|}
 \cline{2-3}
\multicolumn{1}{l|}{}& Non-circular & Circular\\ \cline{1-3}
$1$-way & & $c$-ORing, $c$-Square of ORings\\ \cline{1-3}
$2$-way & $c$-Line, $c$-Square & $c$-Ring, $c$-Square of Rings\\  \cline{1-3}
\end{tabular}
}
\caption{Models of cellular automata.}\label{schema}
\end{figure}

To define the behaviour of all the introduced models, we use the symbol
$Q$ referring to the set of states of a given cellular automaton
$A$. Different  transition functions are defined for different communication complexities.
If we consider $c$-links then for the non-boundary cells
the transition function is $\delta: D_A^c \rightarrow D_A^c$, 
where $D_A^c$ is the set of tuples
$(q,s_1,\cdots ,s_{m_A})$ with $q \in Q$ and $s_j \in \{0,1\}^c $.
In the non-circular models we should also define transition functions
for the boundary cells (recall these
cells are connected to less adjacent cells).
We omit 
the formal definitions of these functions since they are quite standard
and can be easily obtained by the definition for the non-boundary cells.
The behaviour of a cell $i$ can be described as follows.
Let $\delta(q,r_1,\cdots,r_m)=(p,s_1,\cdots,s_m)$,
if a cell $i$ 
is in the state $q$ and receives $r_1,\cdots,r_m$ from the cells
in $N(i)$, then it enters the state $p$ and sends the words
$s_1,\cdots,s_m$. (Note that  this definition is symmetric: the number
of words that each cell sends coincides with the number of
received words.)
 
Note that in the standard definition of cellular automaton each 
cell can send to its neighbouring cells just its state. Therefore, in 
this paper, whenever we consider a model with
link capacity $c$ such that $c \geq \lceil \log |Q| \rceil$, we 
will omit the index $c$ (that is, we will just speak about a Line, Square, 
etc., instead of a $c$-Line, $c$-Square, etc.). 
Some of the results given in this paper hold for all the models,
thus we will speak about a $c-CA$ to mean any of the models above
with link capacity $c$.

A {\em configuration} of a one dimensional cellular automaton with $c$-links
is a mapping $C:\{0,\ldots,n-1\}\rightarrow D_A^c$.
At time $t$, a configuration gives,
for each cell $i$, the state entered and the words of bits
sent at this time.
A {\em starting configuration} is a configuration at time $1$.
In the following we often write ``$(A,C)$'' to denote a 
cellular automaton $A$ starting on a configuration $C$.
We consider the {\em time-unrolling} of $A$, that is a time -space array. 
A pair $(i,t)$ in this array,
with $0< i < n$ and $t\geq 1$, is called a {\em site},
and denotes the cell $i$ at time $t$.
The state of the cell $i$ at time $t$ is denoted
by $\stato(i,t)$  and the words of bits sent to the neighbours
are denoted by $\sx(i,t)$ and $\dx(i,t)$.  
Sometimes, to avoid ambiguities, we will use 
$\stato_A(i,t)$,  $\sx_A(i,t)$ and $\dx_A(i,t)$
to denote the state or the words of bits sent by 
a cell at time $t$ in a fixed cellular automaton $A$.
A site $(i,t)$ is said to be
{\em active} if either it changes its states at the next step,
or sends/receives a words different from 0, that is when one of the following conditions holds: 
\begin {itemize}
\item $\stato(i,t)\neq \stato(i,t+1)$, 
\item either $\sx(i,t) \neq 0 $ or $\dx(i,t)\neq 0$,  
\item there is $i' \in N(i)$ such that 
either $\sx(i',t-1) \neq 0 $ or $\dx(i',t-1)\neq 0$.  
\end {itemize}

 In the two-dimensional
cases a configuration is defined in a natural way and
the time-unrolling consists of triple $(i,j,t)$,
with $0< i,j < n$ and $t\geq 1$, denoting the cell $(i,j)$ at
time $t$. The state of the cell $(i,j)$ at time $t$ is
denoted by $\stato(i,j,t)$.

\smallskip\noindent{\bf The problem.}  Here we introduce a synchronization
problem which generalizes the so called Firing Squad Synchronization Problem
(FSSP). Among the states of the considered cellular automaton, there are three distinguished states:
$G$ the {\em General} state, $L$ the {\em Latent} state,
and $F$ the {\em Firing} state.
The state $L$, also said {\em quiescent} as well, has the property that
if a cell in state $L$ receives all words 0  
from its neighbours
 it remains in the same state and sends the word 0 to its neighbours.
A {\em standard configuration} is a configuration where
the cell $0$ (respectively cell ($0,0$) in the two-dimensional case) 
is in state $G$ and sends a word different from 0  to each neighbour and
all the other cells are in state $L$ and send the word 0.

A {\em synchronization in time  $t(n)$}  is a cellular automaton such 
that, starting from a standard configuration,
all cells enter state $F$ at time $t(n)$
for the first time. 
We will speak about a synchronization of a $c$-Line, 
$c$-Square, etc.  
when the cellular automaton is a $c$-Line, a $c$-Square, etc.
Moreover, 
a cellular automaton which provides a synchronization in time $t(n)$ is
also called  a {\em solution in time $t(n)$} of the FSSP, 
or simply a {\em solution}.

We introduce now two variations of the problem whose solutions are sometimes useful 
to synchronize CA. A {\em Two-End synchronization in time $t(n)$} is
a Line such that at time $t(n)$ all cells enter 
for the first time the state $F$, starting from a configuration which differs from
the standard one because both the cell $0$ and the cell $n-1$
are in the state $G$. A {\em Four-End synchronization in time $t(n)$} is
a Square such that at time $t(n)$ all cells enter
for the first time the state $F$, starting from a configuration 
having the cells $(0,0)$, $(0,n-1)$, $(n-1,0)$, $(n-1,n-1)$ 
in the state $G$ and the other cells in the Latent state.
 
It is simple to see that the synchronizations  
of cellular automata with different communication 
complexity are not unrelated problems. Actually,
a synchronization of a $c$-CA
can be seen as a synchronization of a $c'$-CA  
for every $c' \geq c$. In particular we will often use the following
propositions:
 
\begin{proposition}\label{prop1}
If there is a synchronization of a $1-CA$
in time $t(n)$, then there exists a 
 synchronization of a $c-CA$ in time $t(n)$,  for any $c \geq 1$.
\end{proposition}

Note that in literature 
the time taken by a synchronization is sometimes expressed in terms
of the number of steps, see for example \cite{Cu89,Ko77},
and sometimes with the number of successive configurations,
see for example \cite{Ma87, LNP98}.
In this paper the time is expressed by the number of configurations.

\section{Minimal Time Solutions}\label{mts}
In this section we give tight lower bounds on the time of synchronizations
of $c-CA$ and present the algorithms 
for the synchronization in minimal time. 
 
\subsection{Lower Bounds on the Time of the Synchronizations}\label{lb}


A synchronization of a $c$-Line  requires at least time $2n-1$. Intuitively, 
this is the minimal time for the first cell to wake up all the other cells and 
to get back the message that all the cells have been awakened. Recall
that in a starting configuration each cell, except the first, 
is in a Latent state and the cell $i$ can leave the Latent state not before
than time $i+1$.
Thus  all the cells are awake at time $n$, 
and the first cell gets this information back at time 
$2n-1$. 

As regards the two-dimensional cellular automaton, 
Shinahr~\cite{Sh74} has shown that the minimum time for synchronizing a rectangular array of $m \times n$ cells is $n+m+max(n,m)-2$, but this time reduces to
$2n-1$ in the case of a Square. 
The following lemma summarizes these results.

\begin{lemma}\label{lb1d}\label{lb2d}\label{lb2g}
Every synchronization of a $c$-Line or a $c$-Square has time greater than or equal to $2n-1$.
\end{lemma}

The minimum time to synchronize a Ring or Square of Rings is at least,
as above, the time  required by the first cell to send a message
to all the other cells and to get the information back.

\begin{lemma}\label{lb1d2}
Every synchronization of a $c$-Ring or 
 a $c$-Square of Rings has time greater than or equal to $n+1$. 
\end{lemma}

In the next Lemma we show that time $2n$ is necessary to synchronize 
a $c$-ORing and in Lemma~\ref{lb2d2} we show that 
the minimal time is $3n-1$ for a  $c$-Square of ORings.
 
\begin{lemma}\label{lb1di3}
Every synchronization of a $c$-ORing has time greater than or equal to $2n$. 
\end{lemma}
\begin{proof} 
Assume by contradiction that there exists a synchronization
within time $\bar{t}(n)<2n$ of a  ORing (say $A$) and let  
$B$ be an ORing which differs from $A$ just for the size: $B$ has $2n$ cells 
instead of $n$. 
Since for all $t<n$, $\stato_A(n-1,t)=L$ and $\stato_B(2n-1,t)=L$, then
$\bar{t}(n)\ge n$ and 
$\stato_A(i,t)=\stato_B(i,t)$ for all $0\le i\le n-1$ and $1\le t< n$. 
Observe that the state of the cell $n-1$ at time $n+t$, for $0\le t\le n$,
depends on the states at time $n$ of the following cells:
the cells $n-1$ and $n-2$, when $t=1$,
the states of the cells $n-1$, $n-2$ and $n-3$, when $t=2$,
and in General
on the states of the cells $n-1,\ldots,n-t-1$ for $2<t<n$.
As a consequence, $\stato_A(n-1,t)=\stato_B(n-1,t)$  for $1\le t< 2n$. If
$\bar{t}(n)<2n$, then
at time $\bar{t}(n)$ the cell $n-1$
of both $A$ and $B$ will enter the state $F$. 
Anyway the cell $2n-1$ of $B$ at
time $\bar{t}(n)$ is still in the state $L$, thus we have a contradiction.
\end{proof} 
\qed

\begin{lemma}\label{lb2d2}
Every synchronization of $c$-Square of ORings  has
time greater than or equal to $3n-1$.
\end{lemma}
\begin{proof} 
Assume by contradiction that there exists a synchronization 
$A$  in time
$\bar{t}(n)<3n-1$ of a Square of ORings 
and let  and $B$ be a Square of ORings which differs from $A$ for the number of
cells: $2n \times 2n$ instead of $n \times n$.
Since for all $t<n$ and $0\le i\le n-1$, $\stato_A(i,n-1,t)=\stato_A(n-1,i,t)=L$
and $\stato_B(i,2n-1,t)=\stato_B(2n-1,i,t)=L$, then
$\stato_A(i,j,t)=\stato_B(i,j,t)$ for all $0\le i,j\le n-1$ and $1\le t\le n$.
Furthermore, for both $A$ and $B$ the state of cell $(i,j)$ at time $n$ is $L$
for all cells $(i,j)$ such that $i+j>n-1$.
The state of the cell $(n-1,n-1)$ at time $n+t$, for $0\le t\leq \bar{t}(n)-n$,
depends on the states at time $n$ of
the cells $(n-1-u,n-1-v)$, for $u+v\leq t$.
As a consequence, at time $\bar{t}(n)$ the
cell $(n-1,n-1)$ of both $A$ and $B$ will enter the state $F$.
Anyway since the cell $(2n-1,2n-1)$ of $B$ at
time $\bar{t}(n)$ is still in the state $L$, we have a contradiction.
\end{proof} 
\qed
 
\subsection{Synchronization in Minimal Time for Two-way Communication Networks}\label{ub_two}
 
In this subsection we present the minimal time algorithms for
the synchronization of the models whose connections are two-way links. 
The Proposition~\ref{prop1} allows us to prove the statements only for the case 
$c=1$.
 
Waksman in \cite{Wa66} gave the first solution to 
the problem of synchronizing a 
Line in the minimal time $2n-1$, and Mazoyer, in \cite{Ma96} showed that a
minimal time synchronization exists for a $1$-Line. 
Moreover, Shinahr~\cite{Sh74} has shown the minimal time solution
for a Square. In~\cite{LGP04}, the approach by Shinahr is combined with the 
solution by Mazoyer to obtain a minimal time synchronization of 
a $1$-Square.

\begin{lemma}\label{mazo}
For every link capacity $c\geq 1$, there is a synchronization of a $c$-Line
and of a $c$-Square in time $2n-1$.  
\end{lemma}

The above synchronizations can be used to obtain 
a Two-End synchronization of a Line and a
Four-End synchronization of a Square in time $n$ as shown in the
following lemma.

\begin{figure}[tb]
\centerline{\psfig{file=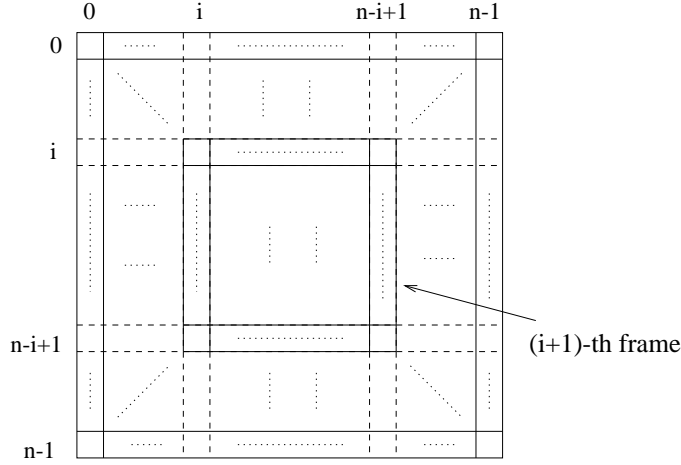,width=9truecm}}
\caption{The frames in a Square of $n\times n$ processors.}\label{frames}
\end{figure}

\begin{lemma}\label{4e}
There are a Two-End synchronization of a Line in time $n$ and
a Four-End synchronization of a Square in time $n$.
\end{lemma}
\begin{proof} 
The Two-End synchronization in time $n$ can be obtained 
by considering a line as split in two halves and synchronizing 
each of them separately by a minimal time solution. 
This can be implemented by just starting a minimal time solution 
from both ends. In fact, each cell can determine its membership to
a sub-line at the time it moves from the Latent state: this happens by
a communication received from its left neighbour (membership to the left 
half-line), or from its right neighbour (membership to the right 
half-line). Note that, in case $n$ is odd, the central cell belongs to 
both half-lines, while when $n$ is even, the central cells 
start acting as the last cells of their half-lines with $1$ time unit of 
delay (at the time they receive a communication from the other half-line). 
Therefore, in both cases the Line is synchronized in time $n$.  

Consider now a Square. We rearrange it
in $n$ concentric frames, where
the $(i+1)$-th inner frame is constituted by the
four lines $(i,i)\ldots(i,n-i-1)$, $(i,n-i-1)\ldots(n-i-1,n-i-1)$,
$(i,i)\ldots(n-i-1,i)$ and $(n-i-1,i)\ldots(n-i-1,n-i-1)$,
see Figure~\ref{frames}.
Suppose now that the cells $(0,0),(0,n-1),(n-1,0)$ and $(n-1,n-1)$ are all
in the same General state.
The four lines of the first frame
can all synchronize in time $n$ using the above result on the 
Two-End synchronization of a Line;
during such synchronizations, after the first two steps,
the four cells $(1,1)$, $(1,n-2)$, $(n-2,1)$ and $(n-2,n-2)$ all
enter a General state and thus the four lines of the second frame
can synchronize in time $n-2$.
Iterating this argument, the $i$-th frame synchronizes
in time $n-2(i-1)$, $1\le i\le \lceil n/2\rceil$.
As this synchronization starts at time $2(i-1)+1$,
then the overall time to synchronize the processors is still $n$. 
\end{proof}
\qed

The synchronizations sketched in the above proof do not work 
when the link capacity is  $1$. The main reason is that 
synchronizations of $1$-CA 
critically use the parity of the time a bit $1$ is received to 
distinguish between different messages.
In particular, 
each cell $i$  expects an even time delay between the 
message sent by the General to wake up all cells and 
the reply sent by the last cell in the Line 
(in a minimal time solution the last cell replies as soon as 
it gets awakened). In the schema sketched in the proof of Lemma~\ref{4e} for 
the Two-End synchronization of a Line, when $n$ is even, the central 
cells delay the response of $1$ time unit. Therefore, the reply message would 
be misunderstood by all the other cells, unless we delay it by another 
time unit. This is the idea exploited in the solution given in 
\cite{LGP04}. Therefore, we have the following lemma.

\begin{lemma}\label{1-4e}
There are a Two-End synchronization of a Line in time $2\lfloor n/2\rfloor +1$ 
and a Four-End synchronization of a Square in time $2 \lfloor n/2 \rfloor +1$.
\end{lemma}

Note that the minimal time synchronization of a $1$-Line 
by Mazoyer \cite{Ma96}, 
can be modified to work for a $2$-Line  without relying on the parity of 
delays to recognize messages (we simply use the second bit to do that).
Thus, it is easy to verify that the schema sketched in the proof of 
Lemma~\ref{4e} can be adapted to work for a $2$-CA using techniques 
similar to that used in \cite{LGP04} for the $1$-CA.
Therefore, the following lemma holds.

\begin{lemma}\label{2-4e}
For every link capacity $c\geq 2$, there are a Two-End synchronization 
of a Line in time $n$ and
a Four-End synchronization of a Square in time $n$.
\end{lemma}

The following lemma states that the lower bounds given in the previous section for $c$-Ring and  $c$-Square of Rings are tight for $c\geq 2$. 
Note that, for the Ring,  a similar, but not correct result, 
can be found in \cite{Cu89}. 

\begin{lemma}\label{ub1dcor}
For every link capacity $c\geq 2$, there is a synchronization of a $c$-Ring and
of a $c$-Square of Rings in time $n+1$. Moreover, 
there is a synchronization of a $1$-Ring and of a $1$-Square of Rings in 
time $2 \lceil n/2\rceil +1$.
\end{lemma}
\begin{proof} 
A $c$-Ring can simulate a Two-End synchronization of a $c$-Line of $n+1$
cells, so obtaining a synchronization in time $n+1$. 
Actually, the cell $0$ can act as both the boundary cells of the $c$-Line.

A synchronization of a $c$-Square of Rings in time $n+1$ can be obtained
by looking at this Square as split in three parts: the first row, the first 
column and the remaining of the array, that is a subarray of $(n-1) \times (n-1)
$ cells.
As we have just noticed, the first row and the first column can be synchronized in time $n+1$.
During these synchronizations (in the first two steps) the cells 
$(1,1)$, $(1,n-1)$, $(n-1,1)$, $(n-1,n-1)$  can enter a new state acting as
a General state of a Four-End synchronization of a Square of $(n-1) \times (n-1)$ cells.
Using Lemmas~\ref{2-4e} and \ref{1-4e} and 
considering that this last synchronization starts 
with a two step delay, we get the stated results.
\end{proof} 
\qed

\bigskip
We can give now the main results of the section.

\begin{theo}\label{basicone}
\qquad

\begin{itemize}
\item  For every link capacity $c\geq 1$, there is a synchronization 
of a $c$-Line and of a 
$c$-Square in time $2n-1$; moreover, every synchronization of a $c$-Line or a $c$-Square has time greater than or equal to $2n-1$.
\item  For every link capacity $c\geq 2$, 
there is a synchronization of a $c$-Ring 
and  of a $c$-Square of Rings in time $n+1$, and 
there is a synchronization of a $1$-Ring and of a $1$-Square of Rings in 
time $2 \lceil n/2\rceil +1$; moreover,
for every link capacity $c\geq 1$, every synchronization of a $c$-Ring 
or a $c$-Square of Rings has time greater than or equal to $n+1$.
\end{itemize}
\end{theo}

We observe that there is a gap between the shown lower and upper bounds 
for the synchronization of a $1$-Ring and a $1$-Square of Rings only for
when $n$ is odd. 

\subsection{Synchronization in Minimal Time for One-way Communication Networks}\label{ub_one}

The following two lemmas state that the lower bounds given in the previous section
for the models using one-way links are  tight.

\begin{lemma}\label{ub1d}
There is a synchronization of a ORing in time $2n$.
\end{lemma}
\begin{proof} 
Using standard techniques, a computation of a Line $A$ of $n$ processors
 in time $t(n)$ 
can be executed by an ORing $B$ in time $2t(n)$, provided that the initial 
configuration of $A$ can be reached  in one step from
the initial configuration of $B$. 
We informally use an induction on the number of steps.
Let $\stato_B(i+1,1)=\stato_A(i,1)$  and  $\stato_B(0,1)=\stato_A(n,1)$ 
and assume that
$\stato_B(i+t,2t)=\stato_A(i,t)$.
(To be more precise, since the cell $i+t$
of $B$ has to simulate the cell $i$ of $A$, then when $i=0$ or 
when $i=n-1$ the state of the cell $i+t$ of $B$ encodes a state of $A$
and the information that the simulated cell is the leftmost or the
rightmost in the line).
Now the cell $i$ of $A$ at the next step needs the states of
cells $i-1$ and $i+1$ at the time $t$. 
Cell $(i-1)+t$ of $B$  passes
its own state $p$ to the cell $(i+t)$ and this in turn forwards $p$
along with its state
to the right neighbouring cell, the cell $(i+1)+t$. 
This last cell can simulate the behaviour of the cell $i$ of $A$ at 
the step $t+1$. Thus $\stato_B(i+t+1,2(t+1))=\stato_A(i,t+1)$.
The overall simulation takes thus a multiplicative delay factor of two.
 
Let us consider now a Two-End synchronization $S$ of a Line.
It takes time $n$ and a synchronization of
an ORing in time $2n$ can be obtained with the 
above simulation. Actually, in the first step it lets the second 
cell enter a General state, so that the state of the cell $i+1$ after the first 
step  is equal to the state of the cell $i$ in the starting
configuration of $S$.
\end{proof} 
\qed

\begin{lemma}\label{ub2d}
There is a synchronization of a Square of ORings in time $3n-1$.
\end{lemma}
\begin{proof} 
We will first give an easier to describe solution which takes time $3n$
and then we show how to save one time unit.

Using standard techniques (as in the previous proof), any computation of 
a Square $A$ in time $t(n)$ can be executed by 
a Square of ORings   $B$ in time $3t(n)$ in the following way.
We informally use an induction on the number of steps.
Assume that the cell $(i+1,k+1)$ in the third configuration of $B$ 
contains the state that the cell $(i,k)$ has in the first configuration 
of $A$ and that cell $(i+j,k+j)$ of $B$ at the time $3j$
has the state that cell $(i,k)$ of $A$ has at the time $j$.
Actually, when the cell $(i,k)$ is a border cell, 
i.e. when either $i \in \{0,n-1\} $ or $k \in \{0,n-1\}$,
also this information is 
stored in the state of the cell $(i+j,k+j)$ of $B$.
Now the cell $(i,k)$ of $A$ at the $j$-th step computes the new state 
from its own state and the states of
cells $(i-1,k)$, $(i,k-1)$, $(i+1,k)$ and $(i,k+1)$ at time $j$.
Within three steps the cell $(i+(j+1),k+(j+1))$ of $B$ can collect the
states that at time $3j$ are in the cells $(i+j,k+j)$,
$((i-1)+j,k+j)$, $(i+j,(k-1)+j)$, $((i+1)+j,k+j)$ and $(i+j,(k+1)+j)$.
Namely:
\begin{enumerate}
\item at step $3j$, cell $(i+j,k+j)$ of $B$ stores the two states $p,q$ of
cells $((i-1)+j,k+j)$ and $(i+j,(k-1)+j)$;
\item at step $3j+1$ the states $p,q$ are passed to cells
$((i+1)+j,k+j)$ and $(i+j,(k+1)+j)$ (note that in the previous step the state
of cell $(i+j,k+j)$ at time $3j$ has been passed to these cells);
\item at step $3j+2$, cell $((i+1)+j,(k+1)+j)$ simulates cell $(i,k)$ of $A$
at step $j$.
\end{enumerate}
 So the state of the cell  $(i+(j+1),k+(j+1))$ of $B$ at time $3j+3$ contains
the state that the cell $(i,k)$ of $B$ has at time $j+1$.
The overall simulation takes thus a multiplicative delay factor of three.
 
Let now $A$ be a Four-End synchronization as in Lemma \ref{4e}.
Recall that in this automaton, the Square is seen as organized in 
concentric frames (see Figure~\ref{frames}) 
which are synchronized at the same time $n$.
We can get a Square of ORings  $A'$ which in
the first two steps reaches a configuration such that
the states of all the cells $(0,0),(0,1),(1,0)$ and $(1,1)$ contain the
General state (recall that the states of the cells
 $(0,0),(0,n-1),(n-1,0)$ and $(n-1,n-1)$ in the starting configuration
of the solution $S$ are all the General state). 
Then $A'$ simulates the solution $A$ within time $3n$.
 
Now let us briefly explain how $A'$ can be modified
to save one step, thus reaching time $3n-1$. 
The first $3n-3$ steps (and thus the first $3n-2$ configurations)
remain unmodified. 
Let us observe what follows:
\begin{enumerate}
\item Each cell of $A$ in the configuration $j$ participates for the
synchronization of the frame which it belongs to;
actually each cell participates either only for a row line or only for a
column line of the frame except for the four corner cells of the frame which
participate for both the lines.
The same holds also for $A'$ in the configurations $3j$ (due to the
mapping between the cells of the configuration $j$ of $A$ and
those of configuration $3j$ of $A'$).
 
\item At time $3j+2$ in $A'$, $1 \le j<n$, a cell $(i+(j+1),k+(j+1))$ is aware
of the states at time $3j$ of the following cells:
\begin{itemize}
\item[a)] $((i-1)+(j+1),(k-1)+(j+1))$, $(i+(j+1),(k-2)+(j+1))$, $(i+(j+1),(k-1)+(j+1))$ and $(i+(j+1),k+(j+1))$;
\item[b)] $((i-1)+(j+1),(k-1)+(j+1))$, $((i-2)+(j+1),k+(j+1))$, $((i-1)+(j+1),k+(j+1))$ and $(i+(j+1),k+(j+1))$.
\end{itemize}
\end{enumerate}
Thus at step $3n-2$, the cell $(i+n,k+n)$ can correctly simulate
either cell $(i,k-1)$ or cell $(i-1,k)$ of $S$ at step $n-1$, hence entering
the Firing state. 
In particular the cell $(i+n,k+n)$ simulates the former
if $(i,k-1)$ participates to the synchronization for a row line, or simulates
the latter, if $(i-1,k)$ participates to the simulation for a column line
(note that at least one of these conditions must hold).
Then, there is a  Square of ORings
inch is a synchronization in time $3n-1$.
\end{proof} 
\qed

\bigskip
We can give now the main results of the section.

\begin{theo}
\quad

\begin{itemize}
\item  There is a synchronization of an ORing in time $2n$ and every synchronization of an ORing has time greater than or equal to $2n$.
\item  There is a synchronization of a Square of ORings in time $3n-1$ and every synchronization of a Square of ORings has time greater than or equal to $3n-1$.
\end{itemize}
\end{theo}

\section{Signals}\label{sig}
The framework of a {\em signal} has been introduced 
in \cite{LNP98} to simplify the design of a $c$-Line.
This innovative definition provides a way to modularize
the design of solutions. 
Informally speaking, a signal is a particular set of cells
that at a given time receives/sends a word different from $0$
from/to the adjacent cells. 
In other words a signal describes 
the information flow in the space-time unrolling of a cellular 
automaton, allowing a modular description of the synchronization process,
that is starting from basic signals we combine different signals
to obtain new ones to describe in a more natural way
the synchronizing algorithms. (Let us note that also in \cite{CC84} and \cite{MT94} the
signals were used, anyway there the intended meaning was different).
The scheme used to present some synchronization algorithms 
in time $t> 2n-1$ for a $c$-Line of $n$ processors is the following:
some signals are designed and composed to obtain an overall 
signal that starts from the leftmost processor
and comes back to it in exactly $(t-2n+1)$ time units; then
a minimal time synchronization starts, thus synchronizing
the $n$ processors in time $t$.

We consider the time unrolling of a $c$-Line $A$
a configuration $C$. Define the time
$t_i^{\max}=\max \{ t | (i,t)$ is active$\}$ and
$t_i^{\min}=\min \{ t | (i,t)$ is active$\}$.
Consider the set of all cells $i$ such that there exists at least
an active site $(i,t)$ of $(A,C)$, for such cells $i$
the set of sites $(i,t_i^{\min})$ 
is called the {\em rear} of $(A,C)$ and the set of sites
$(i,t_i^{\max})$ is the {\em front} of $(A,C)$.
Moreover we say that
$(A,C)$ is {\em \border ed} if there exists a subset of $Q$,
called {\em \border}$(A,C)$
such that for all $i\in\{1,\ldots,n\}$,
$\stato(i,t)\in \border(A,C)$ if and only if $(i,t)$ belongs to the
front of $(A,C)$.
The states in $\border(A,C)$ are called {\em \border} states.
In words, a \border\ state appears for the first time (in the time unrolling
of $A$) on the front of $(A,C)$.

Two active sites $(i_1,t_1),(i_2,t_2)$  are {\em consecutive} if
$t_2=t_1+1$ and $i_2\in \{i_1-1,i_1,i_1+1\}$.
A {\em simple signal} of $(A,C)$ is a subset $S$ of temporally
consecutive sites with the property that
if $(A,C)$ is \border ed, then $(i,t_i^{\max})$ belongs to
$S$.
The union of a finite number of simple signals of a given     
$(A,C)$ is called {\em signal} of $(A,C)$.
A graphical representation of a simple signal $S$
is obtained by drawing a straight line between:\\
(i) every pair of sites $(i,t)\in S$ and $(i,t+1)\in S$ and \\
(ii) every pair of sites $(i,t)\in S$ and $(i+1,t+1)\in S$ (resp.
$(i-1,t+1)\in S$) if $\dx (i,t)=1$ (resp. $\sx(i,t)=1$). \\
A graphical representation of a signal is obtained by the
graphical representation of its simple signals.
The {\em length} of a signal $S$ is
$(t^{\max}-t^{\min}+1)$ where $t^{\max}=\max\{t | (i,t) \in
S, 1 \leq i \leq n\}$ and $t^{\min}=\min\{t | (i,t) \in S, 1 \leq i \leq n\}$.
Sometimes, in the rest of the paper we refer to a signal without
specifying an automata and a starting configuration.

\begin{figure}[htb]
\centerline{\epsfig{figure=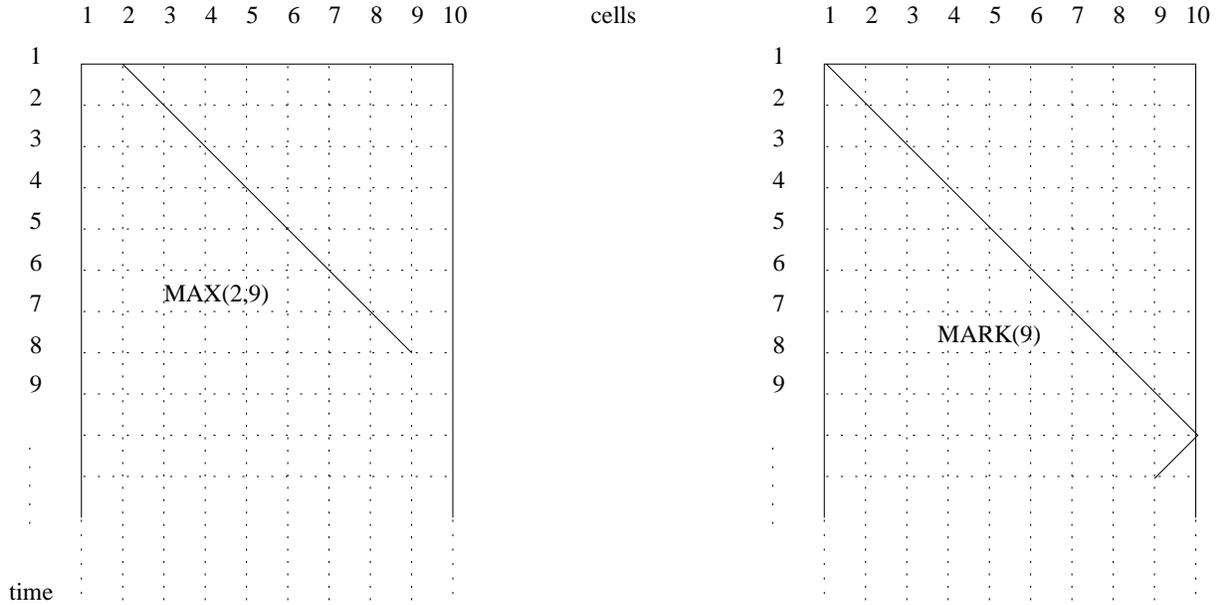, width=16cm}}
\caption{The signals $\Max$ and $\Mark$.}\label{figura1}
\end{figure}

The following examples show two signals: $\Max$ and $\Mark$. The former
is the ``fastest'' signal (it touches one new cell each time unit), while
the latter will be used to 
check the occurrence of an event (generally a signal crossing a given cell)
thus if it is this case, triggering a new signal (see Figure~\ref{figura1}).

\begin{example}\label{exmax}
Let $i\neq j$ and $\Max(i,j)$ be the set containing the sites 
$(i+h,h+1)$
if $i<j$,  or sites $(i-h,h+1)$ otherwise, for $0\leq h\leq |i-j|+1$.
This set is a simple signal,
with length $|i-j|+1$, of a tailed $c$-Line that starts
from a configuration having the states of cells $i$ and $j$ different
from all the others.
\end{example}        
 
\begin{example}\label{exmark}
Given a positive constant $k<n$,
the signal $\Mark(n-k)$ is used to mark the cell $n-k$.
The length of the signal $\Mark$
is $n+k$ (see Figure~\ref{sq}). 
It can be easily seen that $\Mark$ is a signal of a \border ed $c$-Line.
\end{example}         

\subsection{Composition of Signals}
Signals can be composed in order to obtain new ones. 
Given two signals $S_1$ and $S_2$, we define the {\em concatenation}
$\cat_r(S_1,S_2)$ as the signal obtained by starting $S_1$
at time $1$ and $S_2$ at time $r+1$, that is $S_2$ is delayed
$r$ time steps. More formally,
$\cat_r(S_1,S_2)=S_1\cup \{(i,t+r)|(i,t)\in S_2\}$.
In the concatenation of signals the following property is 
crucial.
We say that a $c$-Line $A_2$ on $C_2$ {\em can follow} a \border ed $c$-Line
$A_1$ on $C_1$ if there exists a function $h$ defined over
$\border(A_1,C_1)$
and such that $h(p)=C_2(i)$ if $p=\stato(i,t)$.
When this property holds it is possible to switch from the front 
of $(A_1,C_1)$ to $C_2$.

The following lemma recalls some sufficient conditions
for the existence
of a \border ed $c$-Line for a signal $\cat_r(S_1,S_2)$.
\begin{lemma}\label{sequenziale-bis}\label{cond}
Let $S_1$ and $S_2$ be signals of the \border ed $c$-Lines $(A_1,C_1)$ and
$(A_2,C_2)$, respectively. The signal $S=\cat_r(S_1,S_2)$ is
a signal of a \border ed $c$-Line $(A,C_1)$ if the following conditions
hold:
\begin{enumerate}
\item $(A_2,C_2)$ can follow $(A_1,C_1)$;
\item if a site $(i,t)$ belongs to the front of $(A_1,C_1)$ and
$(i,t')$ belongs to the rear of  $(A_2,C_2)$, then $t< t'+r$;
\item if sites $(i,1)$ and $(j,1)$ belong to the rear of 
$(A_2,C_2)$ then $t_i^{\max}=t_j^{\max}$ in $(A_1,C_1)$.
\end{enumerate}
\end{lemma}
\proof
Let $(i,t)$ be a site of a $c$-Line such that $t$ is the $t_i^{\max}$ in 
$(A_1,C_1)$ and $(i,1)$ belongs to 
the rear of $(A_2,C_2)$. Define $s$ as $r-t+1$.
By the above property $3$, this constant $s$ is well defined, and 
by the above property $2$, it is greater than $0$.
A \border ed $c$-Line $(A,C_1)$ for $S=\cat_r(S_1,S_2)$, 
can be obtained in the following way. 
At the beginning $A$ behaves as $A_1$. 
On the states from $\border(A_1,C_1)$, $A$ counts up to $s-1$ and 
then enter the corresponding state of $C_2$.
We recall that this step is well defined since $s$ is a positive constant and 
the above property $1$ holds. 
At this point $A$ behaves as $A_2$. 
Clearly, $(A,C_1)$ is \border ed and $S$ is a signal of $(A,C_1)$.
Notice that if there are cells corresponding to active sites of $(A_2,C_2)$ which 
do not correspond to active sites of $(A_1,C_1)$, 
from the above properties we have that in both configurations 
$C_1$ and $C_2$ they correspond to quiescent states.
\qed


\subsection{Non trivial signals}\label{el_sig}
We introduce here two non trivial signals of a $c$-Line  that will be used 
to get the main synchronization solutions of the section.
The first has a quadratic length and the second has an exponential length
in the number of cells. In particular from Proposition~\ref{prop1} it is sufficient to consider
only the case $c=1$ (which is also the most difficult).
For technical reasons in this section (and also in section~\ref{ttime}
we will number
the first cell as cell number $1$ (instead of $0$ as said in the preliminaries).

\smallskip\noindent{\bf The signal $\Odd$.} 
Given a positive constant $k < n$, $\Odd(n-k)$ is a signal of a $1$-Line
$A$ which is described as follows:

\begin{itemize} 

\item initially the cell $1$ sends a bit $1$ to the
right;
then if it receives a bit $1$ from the right,
it sends with a delay of one step (except for the first time, when there is 
no waiting), a bit $1$ back
to the right;
the cell $1$ eventually halts when it receives two consecutive bits $1$;  

\item for $1< h< (n-k)$, the cell $h$ sends a bit $1$ to the left 
when it receives for the first time a bit $1$ from the left;
then, if the cell $h$ receives again a bit $1$ from an adjacent cell,
it sends a bit $1$ to the other adjacent cell; 

\item the cell $(n-k)$ sends two consecutive bits $1$ to the left
when it receives a bit $1$ from the left.

\end{itemize} 

\begin{figure}[htb]
\centerline{\epsfig{figure=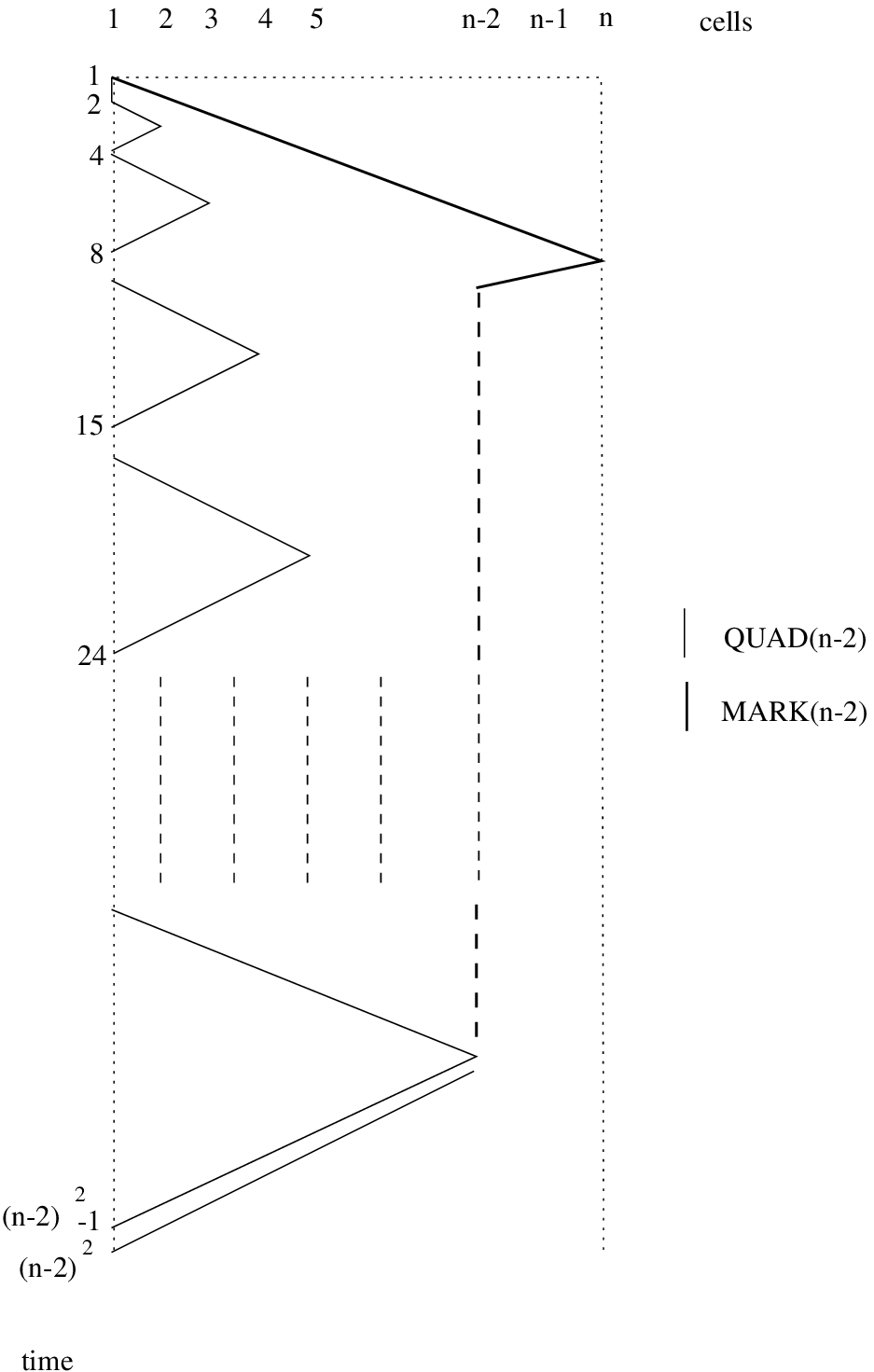, width=8cm}}
\caption{The signal $\cat_1(\Odd(n-2),\Mark(n-2))$.}\label{sq}
\end{figure}

Notice that the designed $1$-Line $A$ can be \border ed as well: in fact the cells from $1$ to $(n-k)$ can enter a \border\ state
when they receive two consecutive bits $1$.
The length of the $\Odd$ signal is $(n-k)^2-1$.  

Let us note now that for the implementation of this signal 
the cell $(n-k)$ needs to be distinguished.
In what follows we will use only $\Odd(n-2)$ in theorem~\ref{bidquadsol} and $\Odd(n-1)$ in 
theorem~\ref{linequadsol}, thus
we only need to distinguish cells $(n-2)$ and $(n-1)$: this can be 
done by $\Mark(n-2)$ and $\Mark(n-1)$, for $n>5$. For smaller
$n$ much easier and ad hoc algorithms can be given
(see Figure~\ref{sq}).

\smallskip\noindent{\bf The signal $\Exp$.} 
Given two positive constants $k$ and $d$,
we will define the signal $\Power(n-k,d)$.

An {\em \stale}
cell is a cell which never sends a bit $1$ unless it receives
a bit $1$ from the left and in this case it sends two consecutive bits $1$ to the left.

Initially the only \stale\ cell is the cell $(n-k)$. 
$\Power(n-k,d)$ is a signal of a $1$-Line which is described as follows:

\begin{itemize}
\item  first cell $1$ sends a bit $1$  
to the right; then, whenever cell $1$ receives a bit $1$ from  
the right, it immediately replies sending back a bit $1$;
finally, if cell $1$ receives two consecutive bits $1$ from the
right, then it changes into an \stale\ cell;

\item for $1 < h < (n-k)$, we distinguish
two cases:
\begin{itemize}
\item if the bit is received from the left then it alternates the following
two behaviours:
\begin{enumerate}
\item it sends a bit $1$ back to the left;  call these
{\em peak cells} (though this is a property of the state entered by this cell.)
\item it sends a bit $1$ to the right; 
\end{enumerate}
each peak cell starts counting from $1$ to $2^{i+1}-2$, for $1< i \leq d$.
When $2^{i+1}-2$ has been just counted,
if the peak cell receives a bit $1$ from the left at the next time unit,
then it is
the $i$-th cell in the line and is marked (see below for an explanation).
This way it can be distinguished later.
\item if a bit $1$ is received from the right, then it sends a bit $1$ to
the left.
If at the next time unit cell $h$ 
receives another bit $1$ from its right neighbour, then two
other sub cases need to be considered:
\begin{itemize}
\item[] if $h > d$ then the cell switches into an \stale\ cell;  
\item[] else, for $h \leq d$, the cell sends two consecutive
bits $1$ to the left.
(Note that when this case occurs, cells $h \leq d$ have already been marked
by step $2$ above.)
\end{itemize}
\end{itemize}
\end{itemize}

\begin{figure}[htb]
\centerline{\epsfig{file=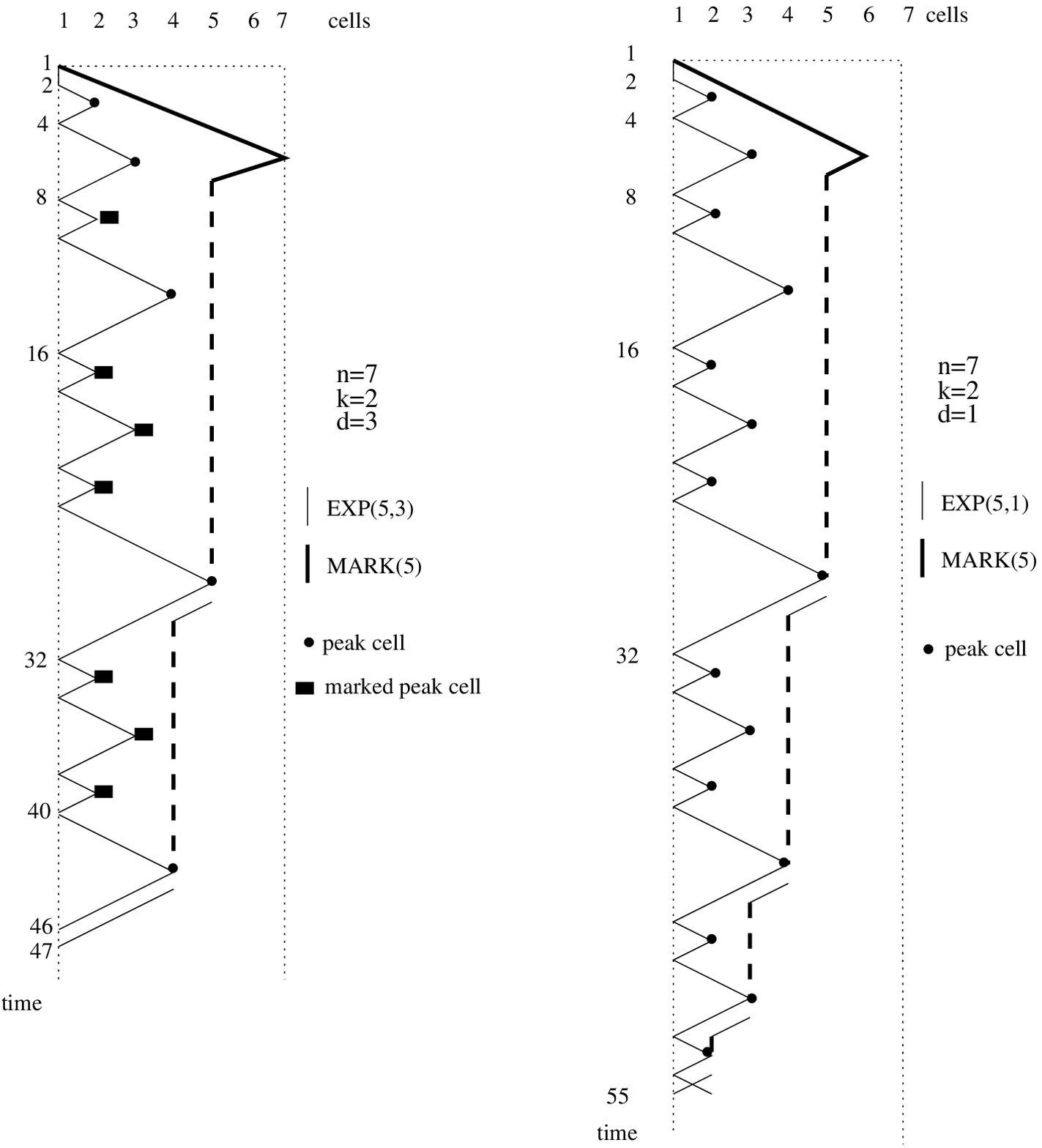, width=12cm}}
\caption{The signals $\cat_1(\Power(5,3), \Mark(5))$ and
$\cat_1(\Power(5,1),\Mark(5))$}\label{exp}
\end{figure}

 From the  algorithm we have just described,
a proof by induction on $i \leq d $ can be given to show how
a peak cell can be marked, in fact the following property holds:
the length of the interval from the instant cell $i$ is a peak cell for the
first time and the instant it becomes a peak cell for the second time
is $2^i+ \sum_{j=1}^{i-1}2^j(i-j)$
(see Figure~\ref{exp} where $d=3$, cell $2$ is marked at time $9$ and 
cell $3$ is marked at time $20$).

\begin{figure}[htb]
\centerline{\epsfig{file=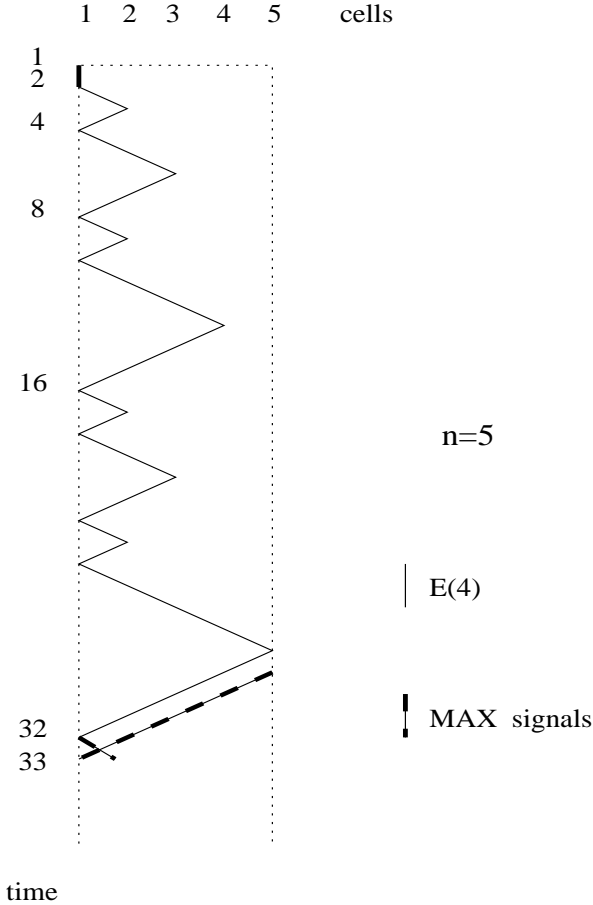, height=12cm, width=8cm}}
\caption{The signals $\Expw(n-k)$}\label{e}
\end{figure}

To implement a \border ed $1$-Line for $\Power(n-k,d)$ initially the
cell $(n-k)$ must be distinguished. In what follows we will
use the signals  $\Power(n-2,\cdot)$ and $\Power(n-1,1)$: the cells
$n-2$ and $n-1$ can be distinguished by using
$\Mark(n-2)$ and $\Mark(n-1)$, for $n>5$. 
Observe also that the cells from $1$ to $(n-k)$ can enter a \border\ state after
they received two consecutive bits $1$.
The length of $\Power(n-k,d)$ is $2^{n-k+1}-2(n-k)-2^{d+1}+2(d+1)$
(see Figure~\ref{exp}).
In a very similar way we can define the signal $\Expw(n-k)$ of
length $2^{n-k+1}+1$ (see Figure~\ref{e}). 

\section{Composition of synchronizations}\label{composition}

The design of synchronizations 
in times which are not minimal may not be obvious.
A compositional approach to achieve this task is thus very 
useful. In this section we discuss several ways to 
combine two or more synchronizations of the models of 
networks we consider. 
We start with a parallel composition, then we study a 
sequential and an iterated compositions.

In the following, if $S_i$ is a synchronization of a $c$-CA
then $G_i$, $L_i$ and $F_i$ denote the General, Latent and
Firing states of $S_i$ and $Q_i$ respectively, $\delta_i$ denote respectively
the set of states and the transition function. 
We use the cross product of automata as a mean to combine $c$-CA.
Given  a $c_1$-Line $A_1$ and a $c_2$-Line $A_2$, 
we denote as  $A_1\times A_2$ the $(c_1+c_2)$-Line
defined as the standard cross product of $A_1$ and $A_2$.
Notice that in the construction we keep distinct the communication links 
of the two lines and thus $A_1\times A_2$ allows to run in parallel
synchronizations of a $c_1$-Line along with synchronizations of a
$c_2$-Line. This construction is extended to all the other 
models we consider in an obvious way. 
We slightly modify the cross product construction to
design a synchronization 
that selects among two different synchronizations 
according to a given condition $P(n)$. 
Examples of such conditions are 
the parity of the number of processors and the fastest/slowest
synchronization.
We define a {\em selecting\/} $c$-Line in time $t(n)$ as a $c$-Line whose state set 
contains two disjoint subsets $O_1$ and $O_2$, called the \emph{selection} subsets, 
such that starting from a standard configuration 
its configuration at any time $t\ge t(n)$ only contains either states from 
$O_1$ or states from $O_2$.
This definition is extended to all the other models we consider in 
an obvious way.
The following lemma shows how to design a $c$-CA 
that selects between two given synchronizations according to a condition 
on the number of cells.
Clearly by iterating this construction, a selection among more than two 
synchronizations can be obtained.

\begin{lemma}\label{pred}
For $i=1,2$, let $S_i$ be a synchronization on a $c_i$-CA
in time $t_i(n)$, and $K$ 
be a selecting $c_K$-CA in time $t(n)\le t_i(n)$ with selection subsets
$O_1$ and $O_2$. Then
there exists a synchronization on a $c'$-CA in time $t'(n)$ 
such that $c'=c_K+c_1+c_2$, moreover
if any configuration of $K$ at time $t\ge t(n)$ contains 
only states from $O_1$ then $s(n)=t_1(n)$, otherwise $s(n)=t_2(n)$.
\end{lemma}
\proof
Let $S$ be the $c'$-CA obtained by modifying 
$K\times S_1 \times S_2$ in the following way:
for $i=1,2$ if a cell is entering $F_i$ and the selecting automaton $K$ is 
in a state from $O_i$ then it enters the firing state of $S$. 
Clearly if $S$ starts on a configuration which is composed of 
triples of corresponding states of the standard configurations 
for $K$, $S_1$ and $S_2$, then $S$ synchronizes in the 
claimed time.
\qed

As applications of the above lemma we show two examples.
In the first example we face with the problem of obtaining a synchronization
which synchronizes at the maximum or at the minimum time
between two synchronizations.
We first define a selecting CA performing the test $t_1(n)\le t_2(n)$,
then we show that this selecting CA can be used to obtain
a synchronization in either the maximum or the minimum time
between two synchronizations.
In the second example a particular behaviour is selected depending
on the result of a comparison between the number of processors $n$
and a constant $h$.

\begin{example}
For $i=1,2$ denote by $S_i$ a synchronization in time
$t_i(n)$. We define a selecting CA $K$ for the condition $t_1(n)\leq t_2(n)$
in time $t(n)=\min\{t_1(n),t_2(n)\}$. 
The CA $K$ is mainly the cross product of $S_1$ and $S_2$ with the modification 
that once a synchronization enters the firing state, $K$ loops on this state. 
Thus we pick $O_1=\{F_1\}$ and $O_2=\{F_2\}$.
Thus by Lemma~\ref{pred} we have 
a synchronization in time $t_1(n)$, if
$t_1(n)\leq t_2(n)$, and $t_2(n)$, otherwise. Thus a synchronization
in the minimum time between $t_1(n)$ and $t_2(n)$ is obtained.
If we pick instead $O_1=\{F_2\}$ and $O_2=\{F_1\}$, then
a synchronization in the maximum time between $t_1(n)$ and $t_2(n)$ 
is obtained.
\end{example}

\begin{example}\label{conta}
We describe a selecting CA $K$ performing the
test $n\leq h$, for a given positive integer $h$.
Let $Q=\{G,L,p_1,\ldots, p_h,$ $p_{\leq h}, p_{>h}\}$ such that $G$ and $L$
are the General and Latent states respectively, and
$O_1=\{p_{\le h}\}$ and $O_2=\{p_{>h}\}$. 
In the linear models the transition function can be 
informally described as follows. 
For the two-dimensional models $K$ can be described in an analogous way.
In the first step cells $0$ and $1$ enter states
$p_1$ and $p_2$ respectively; next, each cell in the Latent state
enters the state $p_{i+1}$ if its adjacent cell on the left is in a
state $p_i$ for $i<h$, while it enters the state $p_{>h}$ if
this neighbour is in the state $p_h$; if each cell is in 
a state $p_i$ for some $i\le h$ thus   
$p_{\le h}$ is propagated up to 
cell $0$ (this takes just a step in a ORing and  $n-1$ steps in a 
Line since this is the case if 
cell $n-1$ is in a $p_i$ for $i\le h$).
When a processor enters the state $p_{\le h}$ or the
state $p_{>h}$ all the other processors are forced
to enter the same state within a time $n$.
Obviously, $K$ is a selecting CA in time $t(n)=n+$min$\{h,n\}$.
\end{example}

Note that the selecting CA from the Example~\ref{conta} can be used
for any pair of synchronizations, as
the time of the selecting CA is not larger than the time of any synchronization.

In the next two lemmas we show how to compose 
two synchronizations in time $t_1(n)$ and $t_2(n)$ respectively,
to obtain
synchronizations in time $t_1(n)+t_2(n)+d$, for a given constant $d$, and in time
$t_1(n) t_2(n)$. 

\begin{lemma}\label{sum}
If $S_1$ and $S_2$ are two synchronizations on a $c$-CA respectively
in time $t_1(n)$ and $t_2(n)$, then there exists a synchronization 
on a $c$-CA in time $t_1(n)+t_2(n)+d$ for $d\ge 0$.
\end{lemma}
\proof
We define a synchronization $S$ such that
$S$ behaves as $S_1$ from time $1$
up to time $t_1(n)$, then at time $t_1(n)+1$ it switches to $S_2$.
Thus $S$ is a synchronization in time $t_1(n)+t_2(n)$.
Furthermore, given a synchronization  $S'$
in time $t(n)$ and with Firing state $F'_0$,
a synchronization in time $t(n)+d$ can be obtained from $S'$
by adding the states $F'_1, \ldots , F'_d$ and the transition rules
from $F'_i$ into $F'_{i+1}$ for $i=0,\ldots ,d-1$, and
picking $F'_d$ as the Firing state of the resulting synchronization.
\qed

\begin{lemma}\label{prod}
If $S_1$ and $S_2$ are two synchronizations on a $c$-CA respectively
in time $t_1(n)$ and $t_2(n)$, then there exists a synchronization 
on a $c$-CA in time $t_{1}(n)\cdot
 t_{2}(n)$.
\end{lemma}

\proof
We prove the above result for a $1$-Line. The proof is similar ofr 
all the other models.
We define  a synchronization $S$ consisting of an Iterative phase
of length $t_1(n)$ which is executed $t_2(n)$ times.
The set of states of $S$ is $Q_1 \times Q_2 \times \{0,1\}^2$,
the General state is $(G_1,G_2,0,1)$, the Latent state is
$(L_1,L_2,0,0)$ and the Firing state is $(F_1,F_2,0,0)$.
In the Iterative phase, the synchronization $S$ modifies the first component
of its state according to the transition functions of $S_1$, until this
component is $F_1$. At the end of this phase $S$ executes a transition
step modifying the
second component of the state according to the transition functions of
$S_2$.
The bits sent according to  transition function of $S_2$ 
are saved in the last two components of each state
according to the order left, right. Moreover, in this same step, $S$ replaces 
$F_1$ with either $G_1$
or $L_1$ (depending on whether the cell is the one triggering in the initial
configuration the firing signal of $S_1$) in the first component.
So the Iterative phase can start again,
until the Firing state is entered by all the cells.
So, the synchronization $S_1$ is iterated exactly $t_2(n)$ times
and $S$ takes time $t_{1}(n) t_{2}(n)$.
\qed

Finally we show a construction that allows to 
obtain synchronizations on a $c$-Square in time 
$t(2n-1)$ provided that there exists a synchronization of 
on a $c$-Line in time $t(n)$.

\begin{lemma}\label{sqtoli}
Given a synchronization on a $c$-Line  in time $t(n)$,
there exists a synchronization on a $c$-Square
in time $t(2n-1)$.
\end{lemma}
\proof
An $(n\times n)$ array can be seen as many lines of $(2n-1)$ cells,
each of them having as endpoints cells $(0,0)$ and $(n-1,n-1)$.
Each of these lines corresponds to a ``path''
from cell $(0,0)$ to cell $(n-1,n-1)$ going through
exactly $(2n-3)$ other cells.
Each cell $(i,j)$ of these paths has as
left neighbour either cell $(i-1,j)$ or cell $(i,j-1)$
and as right neighbour either cell $(i+1,j)$ or cell $(i,j+1)$.

Notice that a cell $(i,j)$ is the $(i+j-1)$-th cell from the left
in all the lines it belongs to. This property allows us to
execute simultaneously on all these lines
a synchronization in time $t(n)$. 
Since the length of each line is $(2n-1)$,
we have a synchronization of $c$-Square in time
$t(2n-1)$.
\qed

\section{Two-way communication Networks}\label{ttime}
In this section we compose the signals presented
in the previous section to obtain solutions in time $n^2$, $2^n$,
$n \lceil \log n \rceil$ and $n \lceil \sqrt n \rceil$ on a $1$-Line and on a $1$-Square.
Clearly these give as a corollary solutions in the same time for the
$c$-Line and $c$-Square and for the circular $c$-Ring and $c$-Square of Rings.

For technical reasons we start numbering cells from $1$ (instead of $0$).
\begin{theo}\label{linequadsol}
There is a synchronization of a $1$-Line in time $n^2$.
\end{theo}
\begin{proof}
The solution is divided into two phases.
The first phase  consists of $\cat_1(\Mark(n-1),\Odd(n-1))$ and
has length $(n-1)^2$ as $\Odd(n-1)$ is delayed one time
step, see  Figure~\ref{sq}.
By Lemma~\ref{sequenziale-bis}, this phase is a signal of a \border ed
$1$-Line starting from a standard configuration.
Hence cell $1$ has entered a \border ed state, say $G'$ and considering
this as the general state, a minimal time solution on a line is started, one step later: this is
the second phase. Together the two phases give a solution to the FSSP in time $n^2$.
\end{proof}
\qed
                            
\begin{theo}\label{bidquadsol}
There is a synchronization of a $1$-Square in time $n^2$.
\end{theo}
\begin{proof}
The algorithm is the following: first a signal
$\cat_1(\Mark(n-2), \Odd(n-2))$ is started on the first row,
the length of this signal is $(n-2)^2$ since $\Odd(n-2)$ is delayed
one time step.
This is a signal of a \border ed $1$-Line starting from a standard
configuration (see Lemma~\ref{sequenziale-bis}).
Thus after $(n-2)^2$ time units the cell $(1,1)$ enters a \border\ state, say $G'$.
Considering $G'$ as the General state, a minimal time synchronization on a
linear array of $n$ cells is executed
on the first row and this takes other $(2n-2)$ time units.
Once the Firing state $F'$ is reached, we use $F'$ as the General
state of a minimal time synchronization that this time runs on each column, thus
taking another $(2n-2)$ time units, which adds up
to a total time of $n^2$.
\end{proof}
\qed

\begin{theo}\label{lineexpsol}
There is a synchronization of a $1$-Line in time $2^n$.
\end{theo}
\begin{proof}
The solution is divided into two phases.
The first phase  consists of $\cat_1(\Mark(n-1),\Exp(n-1,1))$ and
has length $2^n -2n+2$ see  Figure~\ref{exp}.
By Lemma~\ref{sequenziale-bis}, this phase is a signal of a \border ed
$1$-Line starting from a standard configuration.
Hence cell $1$ has entered a \border ed state, say $G'$ and considering
this as the general state, a minimal time solution on a line is started: this is
the second phase. Together the two phases give a solution to the FSSP in time $2^n$.
\end{proof}
\qed

\begin{theo}\label{bidexpsol}
There is a synchronization of a $1$-Square in time $2^n$.
\end{theo}
\begin{proof}
First a signal $\cat_1(\Power(n-2,3),\Mark(n-2))$ 
is started on the first row, see Figure~\ref{exp}.
After $(2^{n-1}-2n-3)$ time units the cell $(1,1)$ enters a \border\
state, say $H$.
This is a signal of a \border ed $1$-Line starting from a standard
configuration (see Lemma~\ref{sequenziale-bis}).
Now the cell $(1,1)$ enters a state $G'$ and
a minimal time synchronization on the first row is
accomplished, using $G'$ as the General state,
thus taking other $(2n-1)$ time units.
Once the Firing state $F'$ is reached, each cell of the first row
enters a state $G''$, and launches the signals $\Mark(n-2)$ and $\Power(n-2,1)$
on each column, using $G''$ as the General state.
This takes another $(2^{n-1}-2n+5)$ time units, which sums up
to time $(2^n-2n+1)$. Finally, a minimal time synchronization on each column is
accomplished, thus reaching time $2^n$.
\end{proof}
\qed

The proof of the existence of a synchronization of a $1$-Line in time
$n \lceil\log n\rceil$ and in time $n \lceil \sqrt n \,\rceil$ is quite
involved and long, see \cite{LNP98}. Here we recall the synchronization
for the $1$-Square.

\begin{figure}[htb]
\centerline{\epsfig{figure=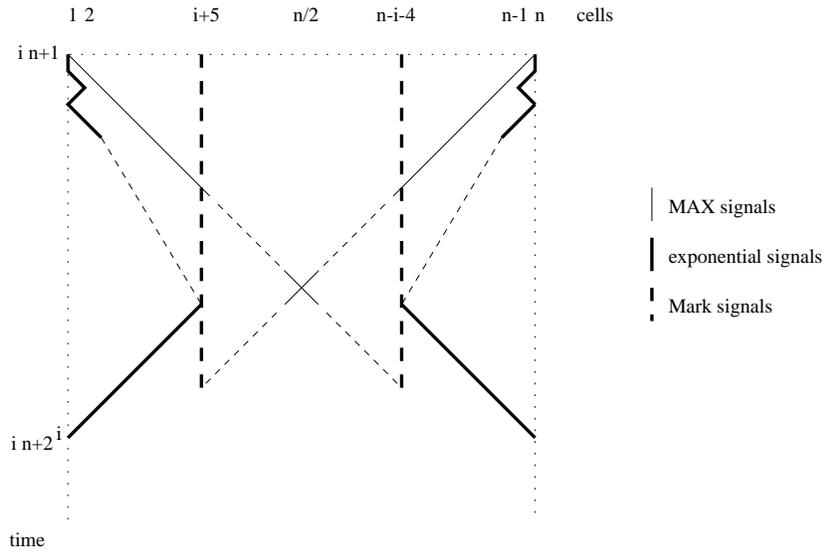, width=11cm}}
\caption{The phase in the i-th iteration,
\mbox{$i>1$} and n odd of the synchronization in time $n \lceil \log n \rceil$.}\label{contb}
\end{figure}

\begin{theo}\label{1squarelogsol}
There is a synchronization of a $1$-Square in time
$n \lceil\log n\rceil$ and in time $n \lceil \sqrt n \,\rceil$.  
\end{theo}
\begin{proof}
The algorithms resemble those used to synchronize a line of $n$ cells at
the same times of \cite{LNP98}. 
Therefore here we only outline the main idea.
For the synchronization in time $n \lceil\log n\rceil$,
 we use a signal to synchronize
the first row in time $(n \log n -2n)$ and then we apply a
synchronization to each column in time $2n$ (just a minimal time synchronization
for a linear array with one more time unit).

Let us informally describe the synchronization of the first row.
Initially the
cells numbered $(1,5)$,$(1,\lceil n/2 \rceil)$, $(1,\lfloor n/2 \rfloor
+1)$
and $(1,n-4)$ are marked: this can be easily accomplished in time $2n$.
This way the row can be seen as split in two halves and for each half a
symmetric computation is done, therefore we will describe only the left half.
A phase is iterated $(\lceil \log n \rceil -5)$ times: each iteration starts
at time
$((i+1) n+1)$, $1 \leq i \leq (\log n -5)$, and has length $n$.
During the $i$-th iteration, the test $(i+5)\ge \lceil \log n \rceil$, is
performed in the following way:
a signal of length $2^{(i+5)}$ on the linear array consisting of the first 
$(i+5)$ cells
and a signal \Max\ of length $n$, which is composed of 
\Max($1,\lceil n/2 \rceil$) and \Max($\lceil n/2 \rceil, 1$), 
are performed (see Figure~\ref{contb}). We compose the
two signals to give \Max\ a higher priority, thus
if the exponential signal reaches a cell after the \Max\ signal, it is
aborted. In this case the \Max\ signal
finishes earlier than or at the same time as the exponential signal,
and this means that $(i+5)\ge\log n$
and thus this is the last iteration.
Otherwise (that is \Max\ finishes
later) cell $(i+1)$ is marked and
a new iteration starts (see Figure~\ref{contb}).
Omitting minor details, at the end of the last iteration all cells are
forced in \border\ states, so determining a standard configuration 
for a 
synchronization of a linear array of $\lceil n/2 \rceil$ cells in time $n$.
The synchronization in time 
$n \lceil \sqrt n \,\rceil$ can be obtained in a very 
similar way by considering a quadratic signal, instead of an exponential one,
to synchronize the first row in time $(n \sqrt n -2n)$. 
\end{proof}
\qed

\begin{corollary}
There are synchronizations of a $c$-Line, $c$-Square and a  $c$-Ring,
$c>1$, in time $n^2$, $2^n$,
$n \lceil\log n\rceil$ and in time $n \lceil \sqrt n \,\rceil$.  
\end{corollary}

\section{One-Way Communication Networks}\label{ttimm}
In this section we give synchronization algorithms for the circular networks, ORings
and Square of ORings, in time $n^2$, $n \lceil \log n \rceil$,
$n \lceil \sqrt n \,\rceil$ and $2^n$.
The algorithms in time $2^n$ is obtained by
converting a solution for a CA.
As in the previous section for technical reasons, 
we start numbering cells from $1$ (instead of $0$).

The following theorem gives the solution in time $n^2$.
\begin{theo}\label{ringquadsol}
There is a synchronization of a ORing and of Square of ORings in time $n^2$.
\end{theo}
\begin{proof}
First consider the ORing.
Assume $n\geq 3$, the case $n<3$ can be dealt with  a simple {\sl ad hoc}
strategy and we omit it (Lemma~\ref{pred} can be used with the test $n\ge3$
to select the behaviour, see Example~\ref{conta}).
The algorithm is very intuitive, thus we will not give the details
of the signals.

The solution is divided into two phases: the {\em Counting} and 
the {\em Synchronization} phases.
The Counting phase has length $(n-2)n+1$ and can be seen as constituted by 
$n-2$ iterations of a sub-phase of $n$ steps. This sub-phase is simply a
$\Max$ signal going all along the ring, from the first cell to the last.
In the first iteration the cell number $3$ is marked with a marker $M$
and at each successive iteration $M$ is moved one cell to the right, so $M$ is 
moved to the first cell when $n-2$ iterations have been executed, that is at 
time $(n-2)n+1$.
This phase is a signal of a \border ed 
$1$-Line starting from a standard configuration (see Lemma~\ref{sequenziale-bis}).             
The Synchronization phase consists of a minimal time solution (in time $2n$) 
on a ring and can start exactly 
at time $(n-2)n+1$ and the total solution has thus length $n^2$.

Now let us consider the Square of ORings.
Here assume $n\ge 5$ and, as before, Lemma~\ref{pred} is used
to select the behaviour.
The solution in time $n^2$ is easily obtained through the
following two steps:
\begin{itemize}
\item the first row is synchronized in time $2n$ with a minimal time
solution on a ring;
\item a solution in time $n^2 -2n$ is applied to each column.
\end{itemize}

The solution in time $n^2 -2n$ is easily obtained from a solution in time 
$n^2$ on a ring and modifying the first iteration of the Counting phase in order to
mark the cell $5$ (instead of cell $3$). In this way the Counting phase is 
constituted by $n-4$ iterations of the sub-phase, thus saving $2n$ steps.
\end{proof}
\qed

Now we show how to obtain a solution in time $n \log n$.
Let us recall that by Lemma~\ref{mazo}
there is a two-end synchronization of a $c$-Line in time $n$.
Notice that it is easy to modify this algorithm in such a way that
in the $(n-1)$-th configuration the processor $\lceil n/2^i\rceil-1$, for
a given $i$, is in a particular state which is different from all other states
entered by the any processor (actually this is a signal of type $\Mark$).
A similar result can be easily obtained for the ORing as well.

\begin{lemma}\label{2end1}
There is a synchronization of a ORing in time $2n$ such that
in the configuration $2n-1$ the processor
$\lceil n/2^i\rceil-1$, for a given $i\ge 0$ is in a particular state which
is different from the state of any other processor.
\end{lemma}

Now we can give the synchronization in time $\lceil n\log n \rceil$.

\begin{theo}\label{torologsol}  
There is a solution of an ORing and of a
Square of ORings in time $n \lceil\log n\rceil$.   
\end{theo}  
\begin{proof}
First consider the ORing.
Let us assume for the moment $n>8$. The solution is divided in three phases:
the {\em Initialization}, the {\em Iterative} and the
{\em Synchronization} phases.
The Iterative phase is executed if $n>16$, otherwise it is skipped.
Informally speaking the whole solution is described as follows.

In the Initialization phase the cell $\lceil n/16\rceil-1$ is marked
with a particular state, call it {\em marker}.
Then the cell $0$ is marked if and only if $n\leq 16$.
Using Lemma~\ref{2end1} this phase can be realized in time $2n$.

In the Iterative phase at the $i$-th iteration the marker is moved from 
the cell $\lceil n/2^{i+3}\rceil-1$ to the cell $\lceil n/2^{i+4}\rceil -1$ for 
$i=1,\ldots ,\lceil\log n\rceil -4$ and again the cell $0$ is marked if 
$n\leq 2^{i+4}$. The $i$-th iteration starts 
at time $(i+1)n+1$ and ends at time $(i+2)n+1$. Note that the  
first step of the $i$-th iteration coincides with the last step of 
the $(i-1)$-th iteration. Thus the total time
taken by this phase is $n(\lceil\log n\rceil -4)+1$.
The third phase is actually a minimal time solution.
Thus, the total time is $2n+n(\lceil\log n\rceil -4)+1+2n-1=n 
\lceil\log n\rceil$. 

The case $n\leq 8$ can be easily solved with a particular 
strategy and the appropriate behaviour can
be selected by using Lemma~\ref{pred}.

Now let us consider the Square of ORings. 
Here assume $n>32$ and, as before, the Lemma~\ref{pred} is used
to choose the behaviour.
The solution in time $n\lceil \log n\rceil$ is easily obtained through the
following two steps:
\begin{itemize}
\item the first row is synchronized in time $2n$ with a minimal time
solution on an ORing;
\item a solution in time $n\lceil\log n\rceil -2n$ is applied to each column.
\end{itemize}

The solution in time $n \lceil\log n\rceil -2n$ is easily obtained from the 
solution in time $n\lceil\log n\rceil$ on a Ring by
modifying the Initialization phase in order to mark the cell 
$\lceil n/64\rceil-1$ (instead of cell $\lceil n/16\rceil-1$) 
thus saving $2n$ steps.
\end{proof}
\qed

\begin{theo}\label{toroexpsol}
There is a synchronization of an ORing and of a 
Square of ORings in time $2^n$.
\end{theo}
\begin{proof}
A synchronization for an ORing in time $2^{n-1}$ can be obtained from Theorem~\ref{lineexpsol}
by putting a General state in the second cell and then
starting a synchronization on $n-1$ cells. In an analogous way it is 
possible to obtain a solution in time $2^{n-2}$.
Using standard techniques as in Lemma~\ref{lb1d}, any computation of a
Line $A$ in time $t(n)$ can be executed by a 
Ring $B$ in time $2t(n)-1$.
In fact, assume that cell $i+j-1$ of $B$ at time $2j-1$ has the
state that cell $i$ of $A$ has at time $j$.
Now the cell $i$ of $A$ at step $j$ needs the states of
cells $i-1$ and $i+1$ at time $j$. Cell $(i-1)+(j-1)$ of $B$ at step $2j-1$ passes
its own state $p$ to the cell $(i+(j-1))$ and this forwards $p$ along with its state
to the right neighbouring cell, the cell $(i+1)+(j-1)$, that at step $2j$ can
simulate cell $i$ of $A$ at step $j$.
Now by this simulation and Theorems~\ref{ringquadsol} and \ref{torologsol}
for the ORing, synchronization algorithms in time
$2^n$ and $2^{n-1}$, respectively, are achieved. 
Moreover, a synchronization of a Square of ORings in time $2^n$ can 
be obtained by first synchronizing the first row in time 
$2^{n-1}$ and then all the columns, with the same algorithm as well.
\end{proof}
\qed

\section{Composed solutions}\label{compo}\label{poly}

In this section we briefly give some new synchronizations on a
$c$-Square using known algorithms to synchronize a
$c$-Line.
Then, we 
show how to construct synchronizations in any time
expressed by polynomials with nonnegative integer coefficients.

In section~\ref{ttime} we have given 
synchronizations for a $c$-Line
in the following times:
$n^2$, $2^n$, $n \lceil \log n \rceil$, and
$n \lceil \sqrt n\, \rceil$.
Combining these results with the Lemma~\ref{sqtoli} 
we can give the following
corollary.
\begin{corollary}
Let $K=2n-1$, there are synchronizations on a $c$-Square in time
$K^2$, $2^{K}$, $K \left\lceil \log K\right\rceil$, and
$K \lceil \sqrt{K}\, \rceil$.
\end{corollary}

The following lemma is crucial to obtain synchronizations 
in polynomial time.
\begin{lemma}\label{poly1}
Given a synchronization on a $c$-CA in time 
$t(n)$ there exist synchronizations in time 
$t(n)+n$ and $n\cdot t(n)$.
\end{lemma}
\begin{proof}
 From Lemma~\ref{mazo}, there exists a 
synchronization on a $c$-Line in time $n$, if the starting configuration 
has the General at both the endpoints.  
We have shown in section~\ref{ub_one} that there exists a 
synchronization on a $c$-Square in time $n$ if the starting configuration 
has the General at all the four corners.
Clearly these synchronizations hold respectively on a $c$-Ring and 
on a $c$-Square of Rings. 
To obtain a synchronization in time $n$ on a $c$-ORing, 
we split the ring in two halves and run the above synchronization 
on a $c$-Line in time $n$ on both the halves at the same time.
This thus requires to start from a configuration where 
the General is at cells $0$, $\frac{n-1}{2}$, $n-1$, if $n$ is 
odd and at cells $0$, $\lfloor\frac{n-1}{2}\rfloor$, $\lceil
\frac{n-1}{2}\rceil$, $n-1$, otherwise.
A synchronization in time $n$ on a $c$-Square of ORings can be 
obtained running the above solution on all the rows at the 
same time and starting from a configuration where the General is for 
$i=0,\ldots,n-1$
at cells $(i,0)$, $(i, \frac{n-1}{2})$,  $(i, n-1)$, if $n$ is
odd and at cells $(i,0)$, $(i, \lfloor\frac{n-1}{2}\rfloor)$,
$(i, \lceil
\frac{n-1}{2}\rceil)$,  $(i, n-1)$, otherwise.
Since on the various models it is possible 
to mark in time $t(n)$ all the cells we need to 
enter the appropriate configuration for the above synchronizations
in time $n$,
we have that by Lemmas~\ref{sum} and \ref{prod} the claimed
synchronizations in time $t(n)+n$ and $n\cdot t(n)$
can be constructed.
\end{proof}
\qed

Thus we have the following theorem.

\begin{theo}\label{psol}
Let $h\geq 2$ be an integer number and $a_0,\ldots,a_h$ be
natural numbers with $a_h \geq 1$.
There is a synchronization in time $a_{h}n^{h}+\ldots+a_{1}n+a_{0}$
on a $c$-Line, a $c$-Square, a $c$-Ring, a $c$-Square of Rings, 
an ORing, and a Square of ORings. 
\end{theo}
\proof
 From 
Corollary~\ref{basicone},
Lemma~\ref{poly1}, and
Theorem~\ref{ringquadsol}, a
synchronization in time $n^b$ can be obtained for every $b\geq 2$.
By composing by Lemma~\ref{sum} these synchronizations in time $n^b$ and 
the minimal time solutions given in sections~\ref{ub_two} and \ref{ub_one}, 
the theorem follows.
\qed

\section{Conclusions}\label{conc}
We have presented various techniques to design solutions
to the FSSP on different kinds of networks. The synchronizing
time is given as input to the problem as is expressed as a function
of the number of nodes.

The approach of the paper has been that of defining a very formal and precise
concept of {\em signal} and starting from basic signals,
give operations to compose them to get other new solutions.
We have introduced also as a parameter the capacity of the link
measured in bits: this has allowed us to classify network models
in terms of the overhead on the amount of traffic on the links.
We believe that this approach can lead to the design of other
signals for new solutions.

Our study has not concerned the problem of the number
of states of the solutions (that in the early papers concerning FSSP
was of primary concerns). As a future direction of research this aspect
has to play a primary role.
Another kind of interesting, but unexplored, question is how to
synchronize a c-line with {\em teratologic} neighbourhoods (for
example (-3,-2,-1,0, 2)), this questions may have some connections with
open questions of~\cite{Ro95}).


\begin{thebibliography}{9999}  
\bibitem{Ba67} R. Balzer, {\em An 8-states minimal time solution to the firing
squad synchronization problem},
Information and Control, 10 (1967), 22--42.

\bibitem{CDDS89} B.A. Coan, D. Dolev, C. Dwork and L. Stockmeyer,
{\em The Distributed Firing Squad Problem},
Siam J. Computing, 18(5), (1989), 990--1012.

\bibitem{CC84} C. Choffrut and K. Culik II, {\em On Real Time Cellular
Automata and Trellis Automata}, Acta Informatica, 21 (1984), 393--407.

\bibitem{Cu89} K. Culik, {\em Variations of the firing squad problem and
applications},
Information Processing Letters, 30 (1989), 153--157.


\bibitem{Go62} E. Goto, {\em A Minimal Time Solution of the Firing Squad
Problem}, {\sl Lecture Notes for Applied Mathematics} 298 (1962), Harvard University,
52--59.

\bibitem{IM96} K. Imai and K. Morita,
{\em Firing squad synchronization problem in reversible cellular automata},
Theoretical Computer Science, 165 (1996), 475-482.

\bibitem{IMS98} K. Imai, K. Morita, K. Sako,
{\em Firing squad synchronization problem in number-conserving
cellular automata},
Proc. of the IFIP Workshop on Cellular Automata, Santiago (Chile), 1998.
                                
\bibitem{Ko77} K. Kobayashi, {\em The Firing Squad Synchronization
Problem for Two Dimensional Arrays}, Information and Control 34 (1977),
153--157.

\bibitem{Ko01} K. Kobayashi, {\em On Time Optimal Solutions of the
Firing Squad Synchronization Problem for Two-Dimensional Paths},
Theoretical Computer Science 259 (2001), 129--143.

\bibitem{LGP04} S. La Torre, J. Gruska, and D. Parente,
{\em Optimal Time\,\&\,{Communication} Solutions of
Firing Squad Synchronization Problems
on Square Arrays, Toruses and Rings},
Proc. of DLT'04,
{\sl Lecture Notes in Computer Science} 3340 (2004), 200--211.
Extended version at URL: {\tt http://www.dia.unisa.it/$\sim$parente/pub/dltExt.ps}

\bibitem{LNP97} S. La Torre, M. Napoli and D. Parente,
{\em Synchronization of 1-Way Connected Processors},
Proc. of the 11th International Symposium on Fundamentals of Computation
Theory, FCT 1997, B.Chelbus and L.Czaja eds.,
Krakow, Poland, September 1 - 3, 1997.
{\sl Lecture Notes in Computer Science} 1279 (1997), 293--304.

\bibitem{LNP98} S. La Torre, M. Napoli and D. Parente,
{\em Synchronization of a Line of Identical Processors at a Given Time},
Fundamenta Informaticae 34 (1998), 103-128.

\bibitem{LNP00} S. La Torre, M. Napoli and D. Parente,
{\em A Compositional Approach to Synchronize Two Dimensional Networks of Processors},
Theoretical Informatics and Applications 34 (2000), 549--564.

\bibitem{Ma87} J. Mazoyer, {\em A six states minimal time solution to
the firing squad synchronization problem},
Theoretical Computer Science 50 (1987), 183--238.

\bibitem{Ma96} J. Mazoyer, {\em On optimal solutions to the firing squad
synchronization problem},
Theoretical Computer Science 168(2) (1996) 367-404.

\bibitem{MT94} J. Mazoyer and V. Terrier, {\em Signals in one dimensional
cellular automata}, Research Report N.94--50, Ecole Normale Superieure
de Lyon, France, 1994.

\bibitem{Mi67} F. Minsky, {\em Computation: Finite and Infinite Machines},
Prentice-Hall, 1967.

\bibitem{Mo64} E. F. Moore, {\sl Sequential Machines, Selected Papers},
(Addison-Wesley, Reading, Mass, 1964).

\bibitem{NH81} Y. Nishitani and N. Honda, {\em The Firing Squad Synchronization
Problem for Graphs}, Theoretical Computer Science 14 (1981), 39--61.

\bibitem{Ro95} Z. Roka, {\em The Firing Squad Synchronization Problem on Cayley
Graphs}, in {\sl Proc. of the 20-th International Symposium on Mathematical
Foundations of
Computer Science MFCS'95}, Prague, Czech Republic, 1995.
{\sl Lecture Notes in Computer Science}, 969 (1995) 402--411.

\bibitem{Sh74} I. Shinahr, {\em Two and Three-Dimensional Firing Squad
Synchronization  Problems}, Information and Control 24 (1974), 163--180.

\bibitem{Wa66} A. Waksman, {\em An optimum solution to the firing squad
synchronization problem},
Information and Control 9 (1966), 66--78.



\end{thebibliography}
\end{document}